\documentclass[aps,prc,preprint]{revtex4-1}
\usepackage[pdftex]{graphicx}
\usepackage{amsmath,amssymb}
\usepackage[makeroom]{cancel}
\usepackage{mathtools}
\usepackage{xcolor}
\usepackage{hyperref}

\hypersetup{
    colorlinks=true,
    linkcolor=blue,
    filecolor=magenta,      
    urlcolor=cyan,
    pdfpagemode=FullScreen,
    }

\begin{document}
\title{Probing Gluon Bose Correlations in  Nuclear Wave Function in Deep Inelastic Scattering}
\date{\today}

\author{Alex~Kovner}
\affiliation{Physics Department, University of Connecticut, 2152 Hillside Road, Storrs, CT 06269, USA} 

\author{Ming~Li}
\affiliation{Department of Physics, The Ohio State University, Columbus, OH 43210, USA}

\author{Vladimir~V.~Skokov}
\affiliation{Department of Physics, North Carolina State University, Raleigh, NC 27695, USA}
\affiliation{RIKEN/BNL Research Center, Brookhaven National Laboratory, Upton, NY 11973}

\begin{abstract}
We extend the results of  [Phys.Rev.Lett. 128 (2022) 18],  where we argued that in the controlled environment of the Deep Inelastic Scattering experiments,  Bose-Einstein correlation between gluons in a hadronic wave function can be accessed through  the production of the   diffractive dijet plus a third jet.  In this observable, Bose-Einstein correlation causes  the enhancement of the production cross sections at the zero relative angle between the transverse momentum imbalance of the photon-going dijet and the transverse momentum of the gluon jet, when the magnitude of the momentum imbalance is about the same as the magnitude of the produced gluon. 
In the present paper, we account for multiple scattering and non-linear effect in the target wave function.  Although our equations can be applied to any high-energy DIS kinematics, to make them tractable numerically, we consider the high-momentum limit (momentum larger than $Q_s$) for the total momentum of the dijet, momentum imbalance, and the momentum of the produced gluon. By performing explicit numerical calculations, we confirm that the signal is present after accounting for  multiple scattering.   
\end{abstract}
\maketitle
\tableofcontents

\section{Introduction}

In the recent letter \cite{Kovner:2021lty} we proposed an observable that directly probes Bose-Einstein correlations between gluons in the nucleus~\cite{Dumitru:2010iy,Altinoluk:2015uaa,Kovner:2018azs}, and is also sensitive to gluon saturation. Study 
of correlations is of course a very exciting subject~\cite{Gaunt:2009re,Blok:2010ge, Diehl:2011yj, Blok:2013bpa,Diehl:2017wew} that goes significantly beyond the single particle distributions traditionally probed in DIS experiments~\cite{Dumitru:2015gaa, Hatta:2016dxp, Dumitru:2018kuw, Mantysaari:2019hkq}. Sensitivity to gluon saturation~\cite{Iancu:2002xk,Gelis:2010nm,Kovchegov:2012mbw} is the most welcome feature as well, as probing gluon saturation is one of the declared aims of the Electron-Ion Collider (EIC) experiments~\cite{AbdulKhalek:2021gbh,Aschenauer:2017jsk}. The observable in question is the correlation between momentum of a gluon jet $\vec p_3$ and the momentum imbalance of a diffractive quark antiquark dijet system $\vec \Delta=\vec p_1+\vec p_2$.  In \cite{Kovner:2021lty} we have studied the signatures of the Bose correlations and have found that they lead to potentially observable enhancement of the jet-dijet cross-section at zero angle between $\vec p_3$ and $\vec \Delta$ in some kinematical regions accessible at EIC.

An intuitive understanding of this enhancement is easiest in the frame where 
 the virtual photon 
 fluctuates into quark-antiquark pair (dipole) which scatters on the gluon field of the fast-moving hadron target. 
The quark and the antiquark are progenitors of the  two  jets with the 
transverse momenta $\vec{p}_{1}$ and  $\vec{p}_{2}$ produced in the final state.
The transverse  momentum imbalance $\vec{\Delta} = \vec{p}_{1} + \vec{p}_{2} $  arises due to transverse momentum transferred to the dijet from the hadron.

Consider now a final state which, in addition to the $q\bar q$ dijet, contains a gluon jet  with transverse momentum $\vec{p}_{3}$ that originates from the hadron. This jet would naturally have  rapidity very different from  that of the other two jets.
Prior to scattering this gluon  in the hadron wave function is Bose correlated 
with an identical gluon (the two have momenta $\vec{k}_{1}  \approx \vec{k}_{2}  $). It is kinematically possible that the momentum imbalance of the produced di jet arises mainly due to the exchange of the gluon  $\vec{k}_{1} $ between the hadron and the dijet,  $|\vec{\Delta}| = |\vec{k}_{1}|$. 
In this situation, the momentum of the produced gluon jet does not change significantly in the scattering ($\vec{p}_{3}  \approx \vec{k}_{2} $) and the primordial Bose-Einstein correlations should lead to the increase in the cross-section of the trijet production, 
when $\vec{p}_{3} \approx  \vec{\Delta} $. A possible peak  at  $\vec{p}_{3} \approx  \vec{\Delta}$ would thus be a clear signature of the gluon Bose enhancement. 
The question is of course, whether the kinematic region in question is large enough so that this enhancement leads to an observable signal. In \cite{Kovner:2021lty} we have studied this question numerically in the dilute approximation treating the target hadron as a dilute object with gluon distribution given by the McLerran-Venugopalan (MV) model~\cite{McLerran:1993ka,McLerran:1993ni}. We have indeed observed a sizable enhancement at like transverse momenta. Interestingly this enhancement became more pronounced once we have introduced a saturation momentum in the target in order to suppress contributions of low transverse momentum gluons. The introduction of the saturation momentum in the framework of dilute approximation is rather {\it ad hoc}, but is nevertheless quite illuminating.

The purpose of the present paper is to analyze the same process beyond the dilute approximation.  We use a somewhat different representation,
in which  the gluon jet is emitted from the dipole and subsequently scatters off the nucleus. If the transverse momentum of the emitted gluon is small; then most of the final gluon momentum will originated from the interaction with the nucleus probing Bose correlations in this way. We stress that the two views of the process are mathematically equivalent which we explicitly show here.  

As in \cite{Kovner:2021lty} we  consider here a trijet configuration where the $q\bar q$ dijet is in the color singlet state (as opposed to Refs.~\cite{Boussarie:2016ogo,Iancu:2021rup} where the rapidity gap is between proton and the trijet) in order to
minimize the effects due to Sudakov radiation~\cite{Sudakov:1954sw,Mueller:2013wwa} . Sudakov radiation from the gluon $\vec{p}_3$ still has to be accounted for, however if the transverse momentum is not too large we do not expect this to qualitatively change the picture. 

The paper is structured as follows. In Sec.~\ref{sec:2} we derive the analytic formulae for the trijet cross-section production on a dense target in terms of the light-like Wilson line operators. At this point our formulae are derived for an arbitrary kinematics of the three jets and in principle can be studied numerically. However the full numerical calculation  would be very complicated. In Sec.~\ref{sec:lm} we therefore consider specifically the kinematical region where the transverse momenta of all jets are much larger than the saturation momentum of the target. We stress that this limit does not simply reduce to the dilute approximation of \cite{Kovner:2021lty} as the presence of the saturation momentum affects the range of the integration over intermediate momenta, and this is important for determining the magnitude of the effect we are after. We find indeed, that consistently with \cite{Kovner:2021lty} the correlated signal is sizable even in this high momentum limit. We conclude with a discussion in Sec.~\ref{concl}.

\section{ Diffractive trijet production in  dense target regime}
\label{sec:2}
We consider trijet production in high energy Deep Inelastic Scatterings. The formalism of our choice is the wave function approach of Refs.~\cite{Kovner:2001vi,Baier:2005dv} (see also Ref.~\cite{Duan:2022pma} where the same approach was applied to two gluon production).  In the rest frame of the nucleus, the virtual photon fluctuates into a pair of quark antiquark, which subsequently radiate one additional gluon before the system of three particles scatters off the nuclear target. We consider the electron/virtual photon moving along positive-$z$ direction while the proton/nucleus along negative-$z$ direction.

The corresponding production cross-section is proportional to  
\begin{equation}\label{observable}
\begin{split}
&\mathcal{O}(p^+_1, \mathbf{p}_1; p_2^+, \mathbf{p}_2;p_3^+, \mathbf{p}_3)\\
=&\sum_{\{c\},\{ \sigma\}} \langle \psi_F | \hat{d}_{c_1,\sigma_1}^{\dagger}(p^+_1,\mathbf{p}_1) \hat{d}_{c_1,\sigma_1}(p^+_1,\mathbf{p}_1) \hat{b}_{c_2,\sigma_2}^{\dagger}(p^+_2,\mathbf{p}_2) \hat{b}_{c_2,\sigma_2}(p^+_2,\mathbf{p}_2) \hat{a}_i^{c_3 \dagger}(p_3^+,\mathbf{p}_3) \hat{a}_i^{c_3}(p_3^+,\mathbf{p}_3)|\psi_F\rangle ,.
\end{split}
\end{equation}
Here $b^\dagger$, $d^\dagger$ and $a^\dagger$ are quark, antiquark and gluon creation operators respectively (and similarly for annihilation operators).
The  final state wavefunction is computed by
\begin{equation}\label{psif}
|\psi_F\rangle = \hat{C}_D^{\dagger} \hat{S}\hat{C}_D |\gamma^{\ast}\rangle\otimes |N\rangle
\end{equation}
with  the initial state wavefunction being the tensor product of the virtual photon state and the nucleus state $|\gamma^{\ast}\rangle\otimes |N\rangle$. 
As usual for high-energy scattering, the virtual photon state $|\gamma^{\ast}\rangle$ is approximated by a quark-antiquark pair. Here $\hat{S}$ is the S-matrix operator.

In the eikonal approximation, the radiation of a gluon from the quark-antiquark dipole is obtained by applying the coherent state operator $\hat{C}_D$~\cite{Kovner:2005pe,Kovner:2005nq} on the quark-antiquark state, with
\begin{equation}
\begin{split}
\hat{C}_D =& \mathrm{exp} \left\{ i\int d^2\mathbf{x} \,\mathcal{B}_i^a(\mathbf{x}) \int_{\Lambda^+ e^{-\Delta y}}^{\Lambda^+} \frac{dk^+}{\sqrt{2\pi} |k^+|} \Big(\hat{a}_i^{a\dagger}(k^+, \mathbf{x}) + \hat{a}_i^a(k^+, \mathbf{x})\Big) \right\}\\
\end{split}
\end{equation}
where $\mathcal{B}_i^a(\mathbf{x})$ is the Weiszacker-Williams (WW) field generated by the dipole. 
The final state radiation off the dipole after interacting with the nucleus is generated by the action of $\hat{C}_D^{\dagger}$ in \eqref{psif}.

For the quark-antiquark dipole, the dilute approximation is well justified; this allows us to simplify the WW field operator and relate it  to the color charge density of the dipole by 
\begin{equation}
\hat{\mathcal{B}}_i^a(\mathbf{x}) = \frac{\partial_i}{\partial^2} \hat{j}^a_D(\mathbf{x}) \equiv \int d^2\mathbf{y} \partial_i \phi(\mathbf{x}-\mathbf{y}) \hat{j}_D^a(\mathbf{y}).
\end{equation}
Here $\partial_i \phi(\mathbf{x}-\mathbf{y})$ is the WW kernel and $\partial^2 \phi(\mathbf{x}) = \delta^{2}(\mathbf{x})$. 
The color charge density operator of the dipole is
\begin{equation}
\hat{j}_D^a(\mathbf{x}) = g\sum_s \int_{k^+} \Big[ \hat{b}^{\dagger}_{h_1, s}(k^+,\mathbf{x}) t^a_{h_1h_2} \hat{b}_{h_2, s}(k^+,\mathbf{x})+ \hat{d}_{h_1, s}(k^+,\mathbf{x}) t^a_{h_1h_2}\hat{d}^{\dagger}_{h_2, s}(k^+,\mathbf{x}) \Big]\,.
\end{equation}
We will use the shorthand notation $\int_{k^+} = \int^{\infty}_0 \frac{dk^+}{
2\pi}$ in the paper. 
The action of $\hat{j}_D^a(\mathbf{x})$ on a general quark-antiquark state is given by  
\begin{equation}\label{eq:jD_on_qqbar}
\begin{split}
\hat{j}_D^a(\mathbf{x}) |q_{a_2}\bar{q}_{a_1}\rangle =& \hat{j}_D^a(\mathbf{x}) \hat{d}^{\dagger}_{a_1, s_1}(k_1^+, \mathbf{x}_1) \hat{b}^{\dagger}_{a_2, s_2}(k_2^+,\mathbf{x}_2) |0\rangle \\
=&\delta^{(2)}(\mathbf{x}-\mathbf{x}_2)g\hat{d}^{\dagger}_{a_1, s_1}(k_1^+, \mathbf{x}_1) \hat{b}^{\dagger}_{h, s_2}(k_2^+, \mathbf{x}) t^a_{ha_2} |0\rangle\\ &-\delta^{(2)}(\mathbf{x}-\mathbf{x}_1)gt^a_{a_1h} \hat{d}^{\dagger}_{h, s_1}(k_1^+, \mathbf{x}) \hat{b}^{\dagger}_{a_2, s_2}(k_2^+, \mathbf{x}_2) |0\rangle \, .
\end{split}
\end{equation}
Acting on a (non-normalized) color singlet dipole state $
|q\bar q\rangle=\sum_a|q_{a}\bar{q}_{a}\rangle $,  one obtains
\begin{equation}
\begin{split}
&\hat{j}_D^a(\mathbf{x}) |q\bar{q}\rangle 
=g\Big[\delta^{(2)}(\mathbf{x}-\mathbf{x}_2) -\delta^{(2)}(\mathbf{x}-\mathbf{x}_1)\Big]t^a_{h_2h_1} \hat{d}^{\dagger}_{h_1, s_1}(k_1^+, \mathbf{x}) \hat{b}^{\dagger}_{h_2, s_2}(k_2^+, \mathbf{x}_2) |0\rangle.
\end{split}
\end{equation}
In order to account for single-gluon radiation, 
we expand the coherent operator $\hat{C}_D$ to the first order in the WW field 
\begin{equation}\label{eq:C_D_to_linear}
\hat{C}_D \simeq 1+ i\int d^2\mathbf{x} \,\hat{\mathcal{B}}_i^a(\mathbf{x}) \int_{\Lambda^+ e^{-\Delta y}}^{\Lambda^+} \frac{dk^+}{\sqrt{2\pi} |k^+|} \Big[\hat{a}_i^{a\dagger}(k^+, \mathbf{x}) + \hat{a}_i^a(k^+, \mathbf{x})\Big]\,.
\end{equation}
Acting on a quark-antiquark state, one obtains the quark-antiquark and gluon Fock state, that is  $|q\bar{q}g\rangle$ 
\begin{equation}
\begin{split}
|q\bar{q}g\rangle =\, & i\int d^2\mathbf{x}  \int_{\Lambda^+ e^{-\Delta y}}^{\Lambda^+} \frac{dk^+}{\sqrt{2\pi} |k^+|} \hat{a}_i^{a\dagger}(k^+, \mathbf{x})\hat{\mathcal{B}}_i^a(\mathbf{x})|q\bar{q}\rangle \\ 
=\, & i g\int_{\Lambda^+ e^{-\Delta y}}^{\Lambda^+} \frac{dk^+}{\sqrt{2\pi} |k^+|} \int d^2\mathbf{x} \Big[\partial_i\phi(\mathbf{x}-\mathbf{x}_2)  t^a_{h_2a_2}\delta_{h_1a_1} -\partial_i\phi(\mathbf{x}-\mathbf{x}_1)t^a_{a_1h_1} \delta_{a_2h_2}\Big]\\
&\qquad \times\hat{a}_i^{a\dagger}(k^+, \mathbf{x})\hat{d}^{\dagger}_{h_1, s_1}(k_1^+, \mathbf{x}_1) \hat{b}^{\dagger}_{h_2, s_2}(k_2^+, \mathbf{x}_2) |0\rangle.
\end{split}
\end{equation}
Again for the color singlet dipole one gets 
\begin{equation}\label{qqbg}
\begin{split}
|q\bar{q}g\rangle =& i g\int_{k^+<\Lambda^+} \frac{dk^+}{\sqrt{2\pi} |k^+|} \int d^2\mathbf{x} \Big[\partial_i\phi(\mathbf{x}-\mathbf{x}_2) -\partial_i\phi(\mathbf{x}-\mathbf{x}_1)\Big] \\
&\qquad \times t^a_{h_2h_1}\hat{a}_i^{a\dagger}(k^+, \mathbf{x})\hat{d}^{\dagger}_{h_1, s_1}(k_1^+, \mathbf{x}_1) \hat{b}^{\dagger}_{h_2, s_2}(k_2^+, \mathbf{x}_2) |0\rangle.\\
\end{split}
\end{equation}
The first term in \eqref{qqbg}  represents to radiation of the gluon from the quark while the second from the antiquark. 
The virtual photon state at NLO in the strong coupling constant $g$ is then approximated by $|\gamma^{\ast} \rangle \sim |q\bar{q}\rangle + |q\bar{q} g\rangle $ \footnote{Strictly speaking one should also keep the $O(\alpha_s)$ virtual correction to the $q\bar q$ component of the state. This $O(\alpha_s$ term however does not contribute in our calculation, and we do not bother writing it explicitly.}.

 In the eikonal approximation,  the S-matrix operator is given by
\begin{equation}
\hat{S} = \mathrm{exp} \left[i \int d^2\mathbf{x} \hat{j}^a(\mathbf{x})\alpha_T^a(\mathbf{x})\right].
\end{equation}
Here the total color charge density $\hat{j}^a(\mathbf{x})$ is a sum of two contributions: one from the quark-antiquark dipole and the other from the emitted gluon $
\hat{j}^a(\mathbf{x}) = \hat{j}^a_D(\mathbf{x}) + \hat{j}^a_G(\mathbf{x}) $
where the latter can be expressed in terms of gluon creation-annihilation operators as  
\begin{equation}
\hat{j}^a_G(\mathbf{x}) = g  \int_{k^+ <\Lambda^+} \frac{dk^+}{2k^+ (2\pi)} \hat{a}_i^{\dagger b}(k^+, \mathbf{x}) T^{a}_{bc}\hat{a}_i^c(k^+, \mathbf{x}).
\end{equation}
Finally, $\alpha_T(\mathbf{x})$ is the classical gluon field of the target which has to be averaged over in the final result with the weight determined by the target wave function.

Explicit calculations lead to the following transformation identities
\begin{equation}
\begin{split}
&\hat{S} \hat{d}^{\dagger}_{h, s}(k^+, \mathbf{x}) \hat{S}^{\dagger} = \mathcal{S}^{\dagger}_{hh'}(\mathbf{x}) \hat{d}^{\dagger}_{h', s}(k^+,\mathbf{x}),\\
&\hat{S} \hat{b}^{\dagger}_{h,s}(k^+, \mathbf{x}) \hat{S}^{\dagger} = \hat{b}^{\dagger}_{h',s}(k^+,\mathbf{x}) \mathcal{S}_{h'h}(\mathbf{x}), \\
&\hat{S}\hat{a}_i^{a\dagger}(k^+,\mathbf{x}) \hat{S}^{\dagger} = \mathcal{U}^{\dagger ab}(\mathbf{x}) \hat{a}_i^{b\dagger}(k^+,\mathbf{x}),\\
\end{split}
\end{equation}
where, on the right-hand side, we have introduced the Wilson line in the { \it foundamental} and the {\it adjoint} representation representations
\begin{equation}
\begin{split}
&\mathcal{S}_{ij}(\mathbf{x}) =\Big[e^{ig\alpha_T^e(\mathbf{x})t^e}\Big]_{ij},\\
&\mathcal{U}^{ ab}(\mathbf{x}) = \Big[e^{igT^e\alpha_T^e(\mathbf{x})}\Big]^{ab}.\\
\end{split}
\end{equation}

The scattering amplitude is obtained by applying $(\hat{S} -1)$ on the virtual photon state at NLO. A simple calculation gives
\begin{equation}\label{eq:S_on_qqbar}
\begin{split}
&(\hat{S} -1)\hat{d}^{\dagger}_{h, s_1}(k_1^+, \mathbf{x}_1) \hat{b}^{\dagger}_{h, s_2}(k_2^+,\mathbf{x}_2) |0\rangle 
=\Big[\mathcal{S}(\mathbf{x}_2)\mathcal{S}^{\dagger}(\mathbf{x}_1) -1\Big]_{h_2h_1} \hat{d}^{\dagger}_{h_1,s_1}(k_1^+,\mathbf{x}_1) \hat{b}^{\dagger}_{h_2, s_2}(k_2^+, \mathbf{x}_2)|0\rangle
\end{split}
\end{equation}
and
\begin{equation}
\begin{split}
(\hat{S}-1) |q\bar{q}g\rangle 
= &i g\int_{k^+<\Lambda^+} \frac{dk^+}{\sqrt{2\pi} |k^+|} \int d^2\mathbf{x} \Big[\partial_i\phi(\mathbf{x}-\mathbf{x}_2) -\partial_i\phi(\mathbf{x}-\mathbf{x}_1)\Big] \\
&\quad \times \Big[\mathcal{U}^{\dagger ab}(\mathbf{x})\mathcal{S}(\mathbf{x}_2) t^a \mathcal{S}^{\dagger}(\mathbf{x}_1)-t^b\Big]_{e_2e_1}\hat{a}_i^{b\dagger}(k^+, \mathbf{x})\hat{d}^{\dagger}_{e_1, s_1}(k_1^+, \mathbf{x}_1) \hat{b}^{\dagger}_{e_2, s_2}(k_2^+, \mathbf{x}_2) |0\rangle\,.
\end{split}
\end{equation}

Finally,  one has to account for  the final state radiation which is generated by the action of the operator $\hat{C}_D^{\dagger}$.  Like in Eq.~\eqref{eq:C_D_to_linear}, one needs to expand the soft gluon coherent operator $\hat{C}_D^{\dagger}$ to the first order in coupling constant.
Since we are working to linear order in $g$, when acting on $(\hat{S}-1)  |q\bar{q}g\rangle$ we can approximate $\hat{C}_D^{\dagger} $ by the unit operator.  The nontrivial contribution comes from applying $\hat{C}_D^{\dagger} $ on $(\hat{S}-1)  |q\bar{q}\rangle$. Using eq.~\eqref{eq:jD_on_qqbar}, it is given by 
\begin{equation}
\begin{split}
&- i\int d^2\mathbf{x}  \int_{\Lambda^+ e^{-\Delta y}}^{\Lambda^+} \frac{dk^+}{\sqrt{2\pi} |k^+|} \hat{a}_i^{a\dagger}(k^+, \mathbf{x})\hat{\mathcal{B}}_i^a(\mathbf{x})\Big[\mathcal{S}(\mathbf{x}_2)\mathcal{S}^{\dagger}(\mathbf{x}_1) -1\Big]_{a_2a_1} \hat{d}^{\dagger}_{a_1,s_1}(k_1^+,\mathbf{x}_1) \hat{b}^{\dagger}_{a_2, s_2}(k_2^+, \mathbf{x}_2)|0\rangle \\ 
=&- i g\int_{\Lambda^+ e^{-\Delta y}}^{\Lambda^+} \frac{dk^+}{\sqrt{2\pi} |k^+|} \int d^2\mathbf{x} \Big[\partial_i\phi(\mathbf{x}-\mathbf{x}_2)  (t^a[\mathcal{S}(\mathbf{x}_2)\mathcal{S}^{\dagger}(\mathbf{x}_1)-1])_{h_2h_1}\\
&\qquad-\partial_i\phi(\mathbf{x}-\mathbf{x}_1)([\mathcal{S}(\mathbf{x}_2)\mathcal{S}^{\dagger}(\mathbf{x}_1)-1] t^a)_{h_2h_1}\Big]\hat{a}_i^{a\dagger}(k^+, \mathbf{x})\hat{d}^{\dagger}_{h_1, s_1}(k_1^+, \mathbf{x}_1) \hat{b}^{\dagger}_{h_2, s_2}(k_2^+, \mathbf{x}_2) |0\rangle\\
\end{split}
\end{equation}

Collecting all the pieces, one obtains the component of the final state wave function that contains quark, antiquark and a gluon. With the leading order virtual photon state 
\begin{equation}
|\gamma^{\ast}\rangle \simeq  \sum_{s_1, s_2} \int_{k_1^+, k_2^+}\int d^2\mathbf{x}_1 d^2\mathbf{x}_2 \Psi^{\gamma^{\ast}\rightarrow q\bar{q}}_{s_1s_2}(k_1^+,\mathbf{x}_1; k_2^+, \mathbf{x}_2) \hat{d}^{\dagger}_{a,s_1}(k_1^+, \mathbf{x}_1)\hat{b}^{\dagger}_{a,s_2}(k_2^+, \mathbf{x}_2) |0\rangle \,,
\end{equation}
we get the relevant component of the final state wave function   
\begin{equation}\label{eq:finalstate_WF}
\begin{split}
&|\psi_F\rangle_{q\bar q g} =   \sum_{s_1, s_2} \int_{k_1^+, k_2^+}\int_{k^+<\Lambda^+} \frac{dk^+}{\sqrt{2\pi} |k^+|}\int d^2\mathbf{x}_1 d^2\mathbf{x}_2 \Psi^{\gamma^{\ast}\rightarrow q\bar{q}}_{s_1s_2}(k_1^+,\mathbf{x}_1; k_2^+, \mathbf{x}_2) 
\\
&\times i g \int d^2\mathbf{x} \Big\{\Big(\partial_i\phi(\mathbf{x}-\mathbf{x}_2) -\partial_i\phi(\mathbf{x}-\mathbf{x}_1)\Big)  \Big[\mathcal{U}^{\dagger ab}(\mathbf{x})\mathcal{S}(\mathbf{x}_2) t^a \mathcal{S}^{\dagger}(\mathbf{x}_1)\Big]_{e_2e_1}\\
 &\qquad-\Big[\partial_i\phi(\mathbf{x}-\mathbf{x}_2)  [t^b\mathcal{S}(\mathbf{x}_2)\mathcal{S}^{\dagger}(\mathbf{x}_1)]-\partial_i\phi(\mathbf{x}-\mathbf{x}_1)[\mathcal{S}(\mathbf{x}_2)\mathcal{S}^{\dagger}(\mathbf{x}_1) t^b]\Big]_{e_2e_1}\Big\}\\
&\qquad \times \hat{a}_i^{b\dagger}(k^+, \mathbf{x})\hat{d}^{\dagger}_{e_1, s_1}(k_1^+, \mathbf{x}_1) \hat{b}^{\dagger}_{e_2, s_2}(k_2^+, \mathbf{x}_2) |0\rangle\,.
\end{split}
\end{equation}

\begin{figure}[t]
\centering 
\includegraphics[width = 0.7\textwidth]{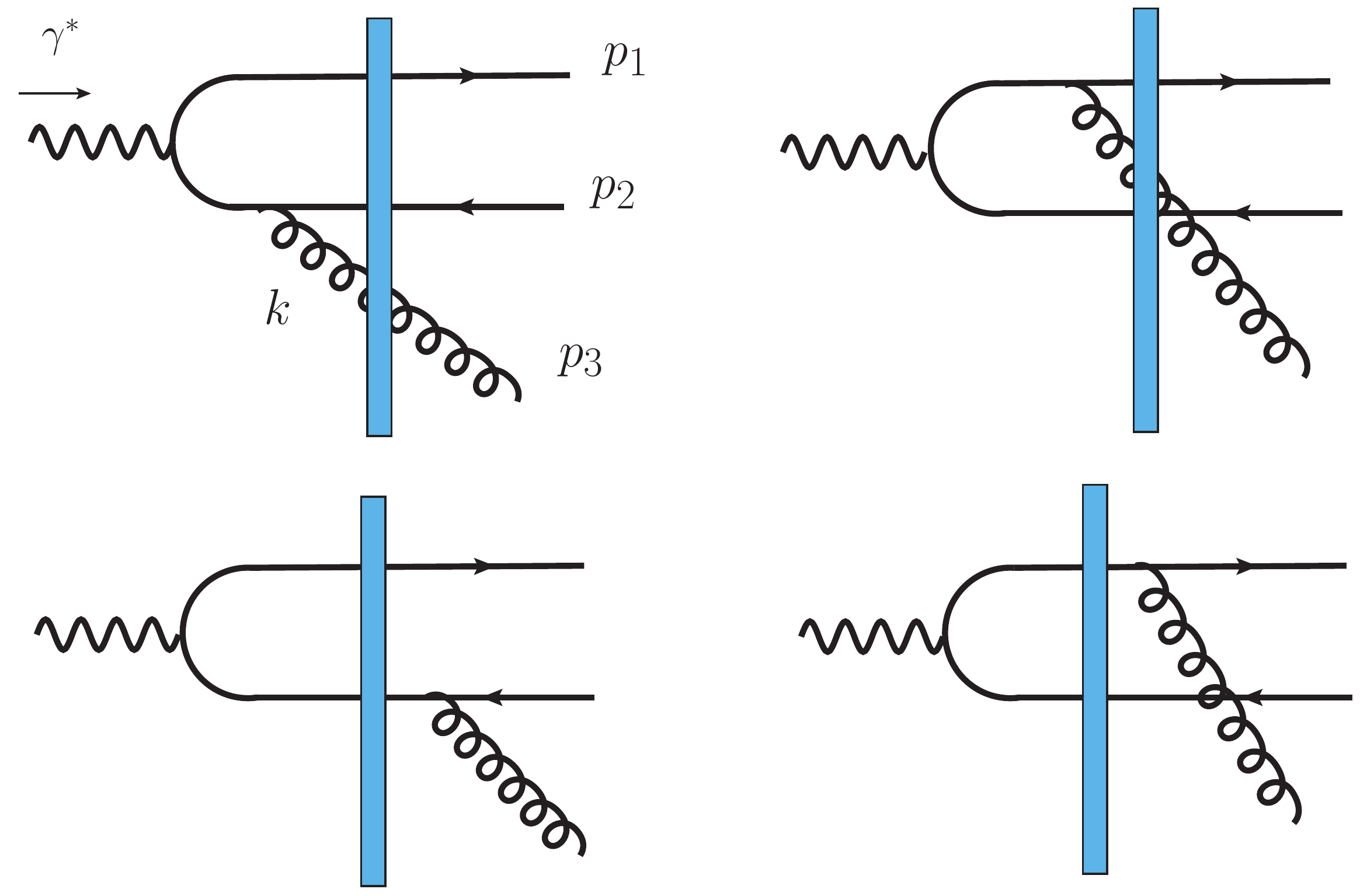}
\caption{Schematic diagram illustrating the trijet production in the electron-going direction. The gluon can be radiated either by the quark or by the antiquark. The radiation can happen either before or after scattering on the target. }
\label{fig:trijet_production_four_amplitudes}
\end{figure}

The observable Eq.~\eqref{observable} in the transverse coordinate space can be expressed as
\begin{equation}\label{eq:observable_mixed}
\begin{split}
&\mathcal{O}(p^+_1, \mathbf{z}_1; p_2^+, \mathbf{z}_2;p_3^+, \mathbf{z}_3)\\
=&\langle \psi_F | \hat{d}_{c_1,\sigma_1}^{\dagger}(p^+_1,\mathbf{z}'_1) \hat{d}_{c_1,\sigma_1}(p^+_1,\mathbf{z}_1) \hat{b}_{c_2,\sigma_2}^{\dagger}(p^+_2,\mathbf{z}'_2) \hat{b}_{c_2,\sigma_2}(p^+_2,\mathbf{z}_2) \hat{a}_j^{c_3 \dagger}(p_3^+,\mathbf{z}'_3) \hat{a}_j^{c_3}(p_3^+,\mathbf{z}_3)|\psi_F\rangle\,\\
=&M(\mathbf{z}_1, \mathbf{z}_2, \mathbf{z}_3)M^{\ast}(\mathbf{z}'_1, \mathbf{z}'_2, \mathbf{z}'_3).\\
\end{split}
\end{equation}
Substituting Eq.~\eqref{eq:finalstate_WF} into Eq.~\eqref{eq:observable_mixed}, 
the trijet production amplitude in coordinate space is given by
\begin{equation}\label{eq:amplitude}
\begin{split}
&M(\mathbf{z}_1, \mathbf{z}_2, \mathbf{z}_3)\\
=&2\sqrt{2\pi} ig \Psi^{\gamma^{\ast}\rightarrow q\bar{q}}_{\sigma_1\sigma_2}(p_1^+, \mathbf{z}_1; p_2^+, \mathbf{z}_2) \Big\{ (\partial_j \phi(\mathbf{z}_3-\mathbf{z}_2) - \partial_j\phi(\mathbf{z}_3-\mathbf{z}_1))\left[\mathcal{U}^{\dagger ac_3} (\mathbf{z}_3)\mathcal{S}(\mathbf{z}_2) t^a \mathcal{S}^{\dagger}(\mathbf{z}_1)\right]_{c_2c_1}\\
&\quad -\left[ \partial_j\phi(\mathbf{z}_3-\mathbf{z}_2)t^{c_3}\mathcal{S}(\mathbf{z}_2)\mathcal{S}^{\dagger}(\mathbf{z}_1) - \partial_j\phi(\mathbf{z}_3-\mathbf{z}_1) \mathcal{S}(\mathbf{z}_2)\mathcal{S}^{\dagger}(\mathbf{z}_1)t^{c_3}\right]_{c_2c_1}\Big\}\,.
\end{split}
\end{equation}
These four terms correspond to the four diagrams in fig.~\ref{fig:trijet_production_four_amplitudes}.

To find the {\it diffractive} dijet production amplitude we need to project $M(\mathbf{z}_1, \mathbf{z}_2, \mathbf{z}_3)$ onto a singlet quark-antiquark dipole state. This amounts to setting $c_1=c_2=c$, summing over the color index $c$ and dividing by the normalization factor $1/\sqrt{N_c}$:
\begin{equation}\label{eq:diffractive_amplitude}
\begin{split}
\mathcal{M}_{\mathrm{diff}} (\mathbf{z}_1, \mathbf{z}_2, \mathbf{z}_3)= &2\sqrt{2\pi} i\frac{g}{\sqrt{N_c}} \Psi^{\gamma^{\ast}\rightarrow q\bar{q}}_{\sigma_1\sigma_2}(p_1^+, \mathbf{z}_1; p_2^+, \mathbf{z}_2) (\partial_j \phi(\mathbf{z}_3-\mathbf{z}_2) - \partial_j\phi(\mathbf{z}_3-\mathbf{z}_1))\\
&\quad \times \Big\{ \mathcal{U}^{\dagger ac_3}(\mathbf{z}_3) \mathrm{Tr}\left[\mathcal{S}(\mathbf{z}_2) t^a \mathcal{S}^{\dagger}(\mathbf{z}_1)\right]-\mathrm{Tr}\left[\mathcal{S}(\mathbf{z}_2)\mathcal{S}^{\dagger}(\mathbf{z}_1)t^{c_3}\right]\Big\}.
\end{split}
\end{equation}

Note that this is the amplitude for the production of a color singlet $q\bar q$ dijet and an additional gluon jet, and not an overall color singlet $q\bar qg$ three jet state.
 The choice of this observable  minimizes the effects of the  Sudakov radiation from the dijet, although as mentioned above it is still susceptible to corrections due to the Sudakov radiation from the gluon jet. In addition, it turns out that in this color configuration,  the back-to-back production of the dijet and the jet is strongly suppressed relative to inclusive states, which makes it a favorable candidate for observation of the same side Bose correlation effect.

Taking the product of the amplitude Eq.~\eqref{eq:diffractive_amplitude} and its complex conjugate, one obtains the final result for diffractive dijet plus gluon jet production 
\begin{equation}\label{trijetc}
\begin{split}
&\mathcal{O}_{\mathrm{diff}}(p^+_1, \mathbf{p}_1; p_2^+, \mathbf{p}_2;p_3^+, \mathbf{p}_3)\\
=&\frac{g^2}{N_c}\sum_{\sigma_1,\sigma_2,j}\int d^2\mathbf{z}_1 d^2\mathbf{z}'_1 e^{i\mathbf{p}_1\cdot(\mathbf{z}_1-\mathbf{z}'_1)} \int d^2\mathbf{z}_2 d^2\mathbf{z}'_2 e^{i\mathbf{p}_2\cdot(\mathbf{z}_2-\mathbf{z}'_2)} \int d^2\mathbf{z}_3 d^2\mathbf{z}'_3 e^{i\mathbf{p}_3\cdot(\mathbf{z}_3-\mathbf{z}'_3)} \\
& \quad \times(8\pi)  \Psi^{\gamma^{\ast}\rightarrow q\bar{q}}_{\sigma_1\sigma_2}(p_1^+, \mathbf{z}_1; p_2^+, \mathbf{z}_2)[\Psi^{\gamma^{\ast}\rightarrow q\bar{q}}_{\sigma_1\sigma_2}(p_1^+, \mathbf{z}'_1; p_2^+, \mathbf{z}'_2) ]^{\ast} \\
&\quad \times \Big[\partial_j \phi(\mathbf{z}'_3-\mathbf{z}'_2) - \partial_j\phi(\mathbf{z}'_3-\mathbf{z}'_1)\Big]\Big[ \partial_j \phi(\mathbf{z}_3-\mathbf{z}_2) - \partial_j\phi(\mathbf{z}_3-\mathbf{z}_1)\Big]\\
&\times\Big\{  [\mathcal{U}^{\dagger}(\mathbf{z}_3)\mathcal{U}(\mathbf{z}'_3)]^{ae} \mathrm{Tr}\left[\mathcal{S}(\mathbf{z}_2) t^a \mathcal{S}^{\dagger}(\mathbf{z}_1)\right]\mathrm{Tr}\left[ \mathcal{S}(\mathbf{z}'_1)t^e\mathcal{S}^{\dagger}(\mathbf{z}'_2) \right]\\
&\qquad -\mathcal{U}^{ c_3e} (\mathbf{z}'_3)\mathrm{Tr}\left[\mathcal{S}(\mathbf{z}_2)\mathcal{S}^{\dagger}(\mathbf{z}_1)t^{c_3}\right]\mathrm{Tr}\left[ \mathcal{S}(\mathbf{z}'_1)t^e\mathcal{S}^{\dagger}(\mathbf{z}'_2) \right]\\
&\qquad -\mathcal{U}^{\dagger ac_3}(\mathbf{z}_3) \mathrm{Tr}\left[\mathcal{S}(\mathbf{z}_2) t^a \mathcal{S}^{\dagger}(\mathbf{z}_1)\right]\mathrm{Tr}\left[ t^{c_3}\mathcal{S}(\mathbf{z}'_1)\mathcal{S}^{\dagger}(\mathbf{z}'_2)\right]\\
&\qquad +\mathrm{Tr}\left[\mathcal{S}(\mathbf{z}_2)\mathcal{S}^{\dagger}(\mathbf{z}_1)t^{c_3}\right]\mathrm{Tr}\left[ t^{c_3}\mathcal{S}(\mathbf{z}'_1)\mathcal{S}^{\dagger}(\mathbf{z}'_2)\right]\Big\}.\\
\end{split}
\end{equation}

A consistency check on this expression is that it reproduces the dilute target limit considered in ref.~\cite{Kovner:2021lty}. To establish this we perform the dilute target expansion in  Appendix A. This exercise has an additional aim to show that the approach taken in ref.~\cite{Kovner:2021lty} where the gluon jet was interpreted as emerging directly from the target, and the approach in the present paper where it is treated as emerging from the perturbative splitting of a dipole, are in fact equivalent and differ only in the choice of the reference frame. 

Returning to our final expression, \eqref{trijetc} we observe that averaging of this expression with respect to target configurations poses a significant numerical challenge. A  simplification is achieved however if we consider a limit of high transverse momenta. This approximation can be implemented analytically up to a point where numerical calculations become feasible. This is what we do in the next section.


\section{Trijet production in high momentum limit}
\label{sec:lm}
\begin{figure}[h]
\centering 
\includegraphics[width = 0.55\textwidth]{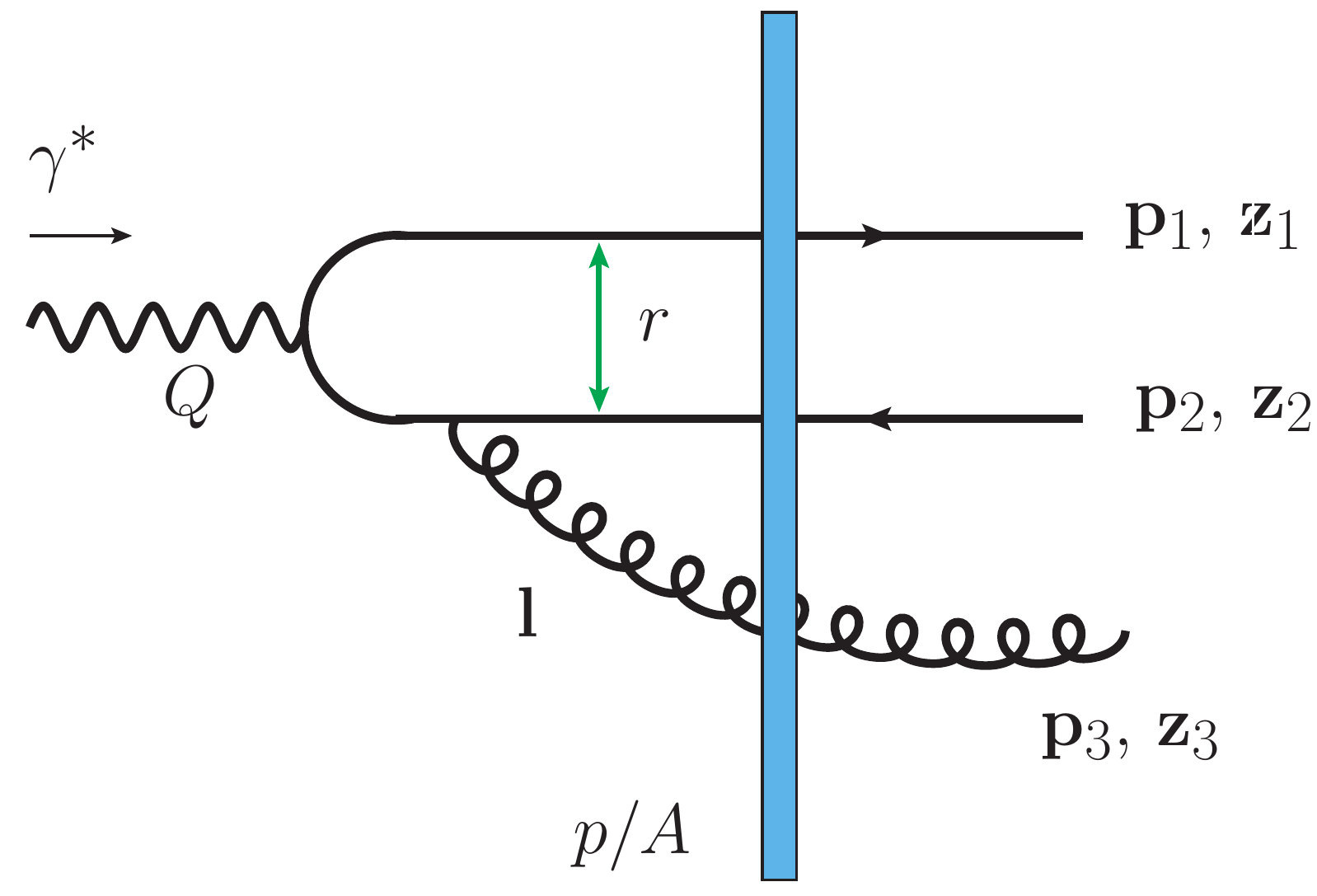}
\caption{Schematic diagram of trijet production in $\gamma^{\ast}+p/A$ scattering. }
\label{fig:LML_illustration2}
\end{figure}

In order to make numerics viable, in this section, we consider the diffractive trijet production in a particular kinematic region: the external momenta $\mathbf{p}_1, \mathbf{p}_2, \mathbf{p}_3$ are taken to be large compared to the saturation scale of the nucleus, $Q_s$. For clarity, the coordinates and momenta of different particles are depicted in fig.~\ref{fig:LML_illustration2}.   

\begin{figure}[t!]
\centering 
\includegraphics[width = 0.5\textwidth]{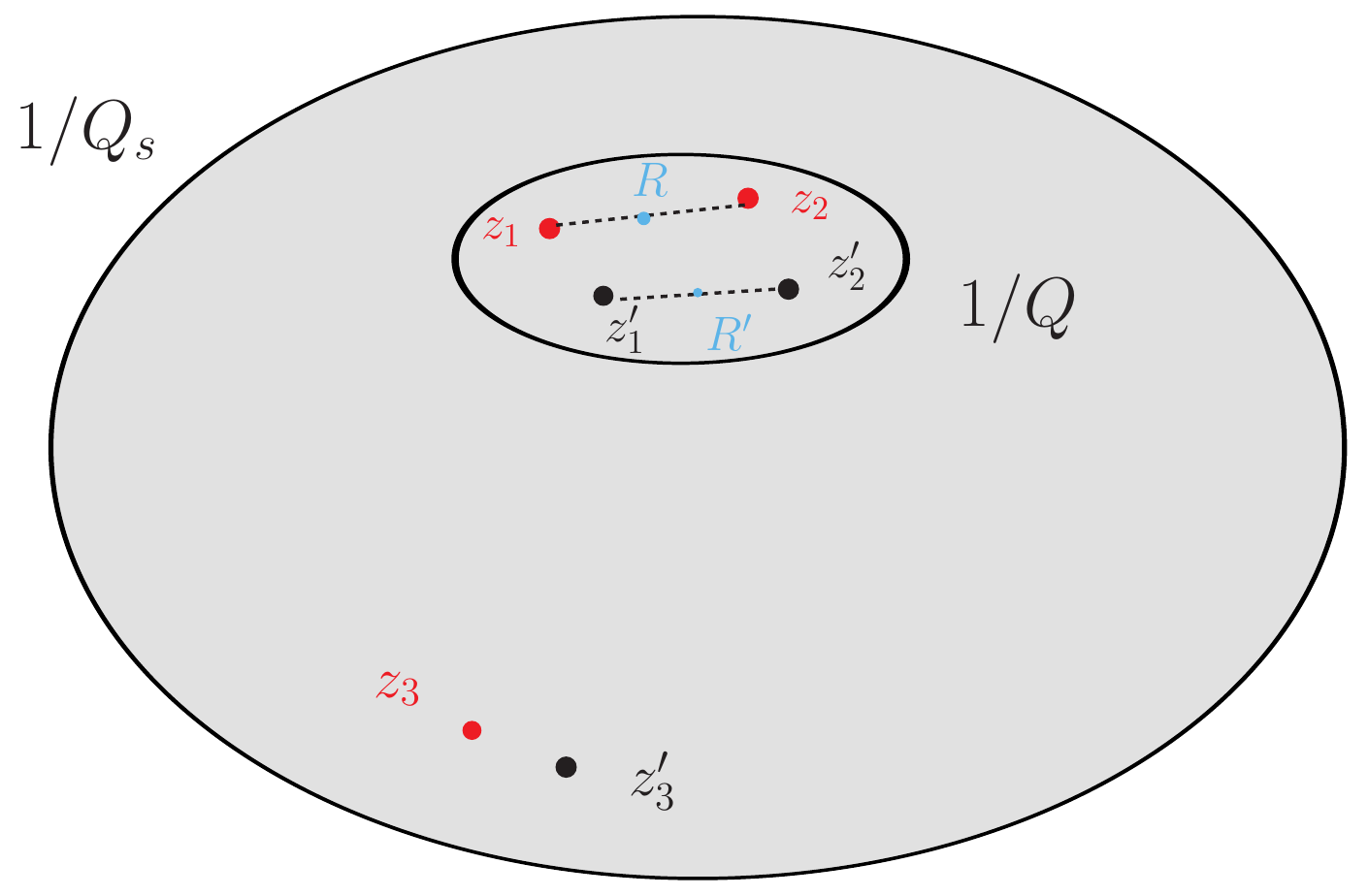}
\caption{A typical coordinate space configuration corresponding to high momentum limit. When $|\mathbf{p}_1|, |\mathbf{p}_2|, |\mathbf{p}_3| \sim Q \gg Q_s$, the coordinates in the amplitude and complex conjugate amplitudes are close within a domain of size $1/Q^2$. The gluon jet is typically separated from the quark and antiquark jets although all of them are expected to be located within the domain of size $1/Q_s^2$. }
\label{fig:trijet_large_momentum_config}
\end{figure}

In the high momentum limit, the coordinates of each particle in the amplitude and conjugate amplitude must be close to each other. Additionally, for high $Q^2$, the quark and antiquark should be close to each other in the coordinate space, as we consider the regime $Q\gg Q_s$.
A typical coordinate space configuration corresponding to the high momentum limit is shown in fig.~\ref{fig:trijet_large_momentum_config}.  


\subsection{High momentum approximation}
We start with the diffractive amplitude for the trijet production \eqref{eq:diffractive_amplitude} obtained in the previous section. 
Using the identity $\mathcal{S}(\mathbf{z})t^a\mathcal{S}^{\dagger}(\mathbf{z}) = \mathcal{U}^{\dagger ab}(\mathbf{z}) t^b$,
one can perform the following rearrangement 
\begin{equation}
 \mathrm{Tr}\left[\mathcal{S}(\mathbf{z}_2) t^a \mathcal{S}^{\dagger}(\mathbf{z}_1)\right] = U^{\dagger ad}(\mathbf{z}_2)\mathrm{Tr}\left[ t^d \mathcal{S}(\mathbf{z}_2) \mathcal{S}^{\dagger}(\mathbf{z}_1)\right] = U^{\dagger ad}(\mathbf{z}_1) \mathrm{Tr}\left[\mathcal{S}(\mathbf{z}_2) \mathcal{S}^{\dagger}(\mathbf{z}_1)t^d\right].
\end{equation}
As a result, the factor involving Wilson lines can be equivalently written as
\begin{equation}
\begin{split}
&\mathcal{U}^{\dagger ac}(\mathbf{z}_3) \mathrm{Tr}\left[\mathcal{S}(\mathbf{z}_2) t^a \mathcal{S}^{\dagger}(\mathbf{z}_1)\right]-\mathrm{Tr}\left[\mathcal{S}(\mathbf{z}_2)\mathcal{S}^{\dagger}(\mathbf{z}_1)t^{c}\right]\\
=&\left(\frac{1}{2}\left[U(\mathbf{z}_3) (U^{\dagger}(\mathbf{z}_2) + U^{\dagger}(\mathbf{z}_1) )\right]^{cd} - \delta^{cd}\right)\mathrm{Tr}\left[\mathcal{S}(\mathbf{z}_2)\mathcal{S}^{\dagger}(\mathbf{z}_1)t^{d}\right].
\end{split}
\end{equation}
For large momenta and large photon virtuality $Q\gg Q_s$, the size of the dipole $|\mathbf{z}_1-\mathbf{z}_2|\sim 1/Q \ll 1/Q_s$. The gluon is typically emitted at larger transverse distances. At the same time, to be sensitive to the correlations in the target, the three particles (quark, antiquark and gluon) have to be located within a transverse size determined by $1/Q_s$. Given these considerations we can perform gradient expansion in various correlators.

To preserve the symmetry between the quark and the antiquark  we find it convenient to introduce the center of mass and the relative coordinates via 
$\mathbf{R} = \frac{1}{2} (\mathbf{z}_2+\mathbf{z}_1)$, $ \mathbf{r} = \mathbf{z}_2-\mathbf{z}_1$. 
Generically at large momenta  $|\mathbf{r}|\ll |\mathbf{R}|$, and we can approximate
\begin{equation}
\mathrm{Tr}\left[\mathcal{S}(\mathbf{z}_2)\mathcal{S}^{\dagger}(\mathbf{z}_1)t^d\right]
\approx \mathrm{Tr} \left[\frac{\mathbf{r}^i}{2} \partial^i\mathcal{S}(\mathbf{R}) \mathcal{S}^{\dagger}(\mathbf{R})t^d\right] - \mathrm{Tr}\left[\mathcal{S}(\mathbf{R}) \frac{\mathbf{r}^i}{2} \partial^i \mathcal{S}^{\dagger}(\mathbf{R})t^d\right] =\frac{1}{2} ig \mathbf{r}^i  A^i_d(\mathbf{R})\,,
\end{equation}
where $A^i_a(\mathbf{R})$ is the WW field of the target
\begin{equation}
A^i_a(\mathbf{R})=\frac{2}{ig}\mathrm{Tr}\left[t^a\partial^i \mathcal{S}(\mathbf{R})\mathcal{S}^{\dagger}(\mathbf{R}) \right]\,.
\end{equation}
Additionally, one can approximate 
\begin{equation}
\frac{1}{2}\left[U(\mathbf{z}_3) (U^{\dagger}(\mathbf{z}_2) + U^{\dagger}(\mathbf{z}_1) )\right]^{cd}  =U(\mathbf{z}_3)U^{\dagger}(\mathbf{R}) + \mathcal{O}(\mathbf{r}^2).
\end{equation}
With the above simplifications, the diffractive amplitude becomes
\begin{equation}\label{eq:highp1p2}
\begin{split}
\mathcal{M}_{\mathrm{diff}} (\mathbf{z}_1, \mathbf{z}_2, \mathbf{z}_3)= &2\sqrt{2\pi} i\frac{g}{\sqrt{N_c}} \Psi_{\sigma_1\sigma_2}^{\gamma^{\ast}\rightarrow q\bar{q}}(p_1^+, \mathbf{z}_1; p_2^+, \mathbf{z}_2) \Big(\partial_j \phi(\mathbf{z}_3-\mathbf{z}_2) - \partial_j\phi(\mathbf{z}_3-\mathbf{z}_1)\Big)\\
&\quad \times \Big[U(\mathbf{z}_3)U^{\dagger}(R) - 1\Big]^{cd} \frac{1}{2} ig \mathbf{r}^i  A^i_d(\mathbf{R}).
\end{split}
\end{equation}

Performing Fourier transformations, the production amplitude in momentum space can be written as 
\begin{equation}
\begin{split}
\mathcal{M}_{\mathrm{diff}} (\mathbf{p}_1, \mathbf{p}_2, \mathbf{p}_3) = &\int d^2\mathbf{z}_1 d^2\mathbf{z}_2 d^2\mathbf{z}_3 e^{i\mathbf{p}_1\cdot\mathbf{z}_1}e^{i\mathbf{p}_2\cdot\mathbf{z}_2}e^{i\mathbf{p}_3\cdot\mathbf{z}_3}\mathcal{M}_{\mathrm{diff}} (\mathbf{z}_1, \mathbf{z}_2, \mathbf{z}_3)\\
=&\int \frac{d^2\mathbf{l}}{(2\pi)^2} \mathcal{M}_{\mathrm{dipole}} (\mathbf{P}_{\perp}, \mathbf{l})\mathcal{M}_{\mathrm{nucleus}} (\mathbf{\Delta}, \mathbf{p}_3, \mathbf{l}).\\
\end{split}
\end{equation}  
Here $\mathbf{\Delta} = \mathbf{p}_1+\mathbf{p}_2$ is the quark--anti-quark momentum imbalance  and $\mathbf{P}_{\perp} = \frac{1}{2} (\mathbf{p}_2-\mathbf{p}_1)$ is the total momentum of the pair. 

The dipole  matrix element is
\begin{equation}\label{eq:dipole_amplitude}
\begin{split}
\mathcal{M}_{\mathrm{dipole}} (\mathbf{P}_{\perp}, \mathbf{l}) = &2\sqrt{2\pi} i\frac{g}{\sqrt{N_c}}\int d^2\mathbf{r} e^{i\mathbf{P}_{\perp} \cdot\mathbf{r}} \Big[e^{i\mathbf{l}\cdot\mathbf{r}/2}-e^{-i\mathbf{l}\cdot\mathbf{r}/2}\Big] \mathbf{r}^i \Psi^{\gamma^{\ast}\rightarrow q\bar{q}}(p_1^+,  p_2^+, \mathbf{r}) \frac{i\mathbf{l}_j}{l^2}\\
=&2\sqrt{2\pi} i\frac{g}{\sqrt{N_c}} (-i\partial_{\mathbf{P}_{\perp}^i}) \Big[ \Psi_{\sigma_1\sigma_2}^{\gamma^{\ast}\rightarrow q\bar{q}}(p_1^+, p^+_2, \mathbf{P}_{\perp} + \mathbf{l}/2) - \Psi^{\gamma^{\ast}\rightarrow q\bar{q}}_{\sigma_1\sigma_2}(p_1^+, p^+_2, \mathbf{P}_{\perp} - \mathbf{l}/2)\Big]  \frac{i\mathbf{l}_j}{l^2}
\end{split}
\end{equation}
and  the nucleus matrix element is
\begin{equation}\label{eq:nucleus_amplitude}
\mathcal{M}_{\mathrm{nucleus}} (\mathbf{\Delta}, \mathbf{p}_3, \mathbf{l}) 
=\frac{1}{2} ig \int d^2\mathbf{R} d^2\mathbf{z}_3 e^{i(\mathbf{\Delta}+\mathbf{l})\cdot\mathbf{R}} e^{i(\mathbf{p}_3-\mathbf{l}) \cdot\mathbf{z}_3}  A^i_d(\mathbf{R}) \left[U(\mathbf{z}_3)U^{\dagger}(\mathbf{R})-1\right]^{cd}.
\end{equation}

From the explicit expressions in Eqs.~\eqref{eq:dipole_amplitude} and ~\eqref{eq:nucleus_amplitude}  we obtain the simplified result for diffractive trijet production in the high momentum limit
\begin{equation}\label{eq:amplitude_squared}
\begin{split}
&|\mathcal{M}_{\mathrm{diff}} (\mathbf{p}_1, \mathbf{p}_2, \mathbf{p}_3)|^2  =(2\pi)4\frac{g^2}{N_c}\int \frac{d^2\mathbf{l}}{(2\pi)^2}\frac{d^2\mathbf{l}'}{(2\pi)^2} \frac{\mathbf{l}\cdot\mathbf{l}'}{l^2l^{\prime 2}}\\
&\times \partial_{\mathbf{P}_{\perp}^i}\Big[ \Psi^{\gamma^{\ast}\rightarrow q\bar{q}}_{\sigma_1\sigma_2}(p_1^+, p^+_2, \mathbf{P}_{\perp} + \mathbf{l}/2) - \Psi^{\gamma^{\ast}\rightarrow q\bar{q}}_{\sigma_1\sigma_2}(p_1^+, p^+_2, \mathbf{P}_{\perp} - \mathbf{l}/2)\Big]\\
&\times \partial_{\mathbf{P}_{\perp}^{i'}}\Big[ \Psi^{\gamma^{\ast}\rightarrow q\bar{q}}_{\sigma_1\sigma_2}(p_1^+, p^+_2, \mathbf{P}_{\perp} + \mathbf{l}'/2) - \Psi^{\gamma^{\ast}\rightarrow q\bar{q}}_{\sigma_1\sigma_2}(p_1^+, p^+_2, \mathbf{P}_{\perp} - \mathbf{l}'/2)\Big]^{\ast}\\
&\times \mathcal{M}_{\rm{nucleus}}(\mathbf{\Delta}, \mathbf{p}_3, \mathbf{l})\mathcal{M}^{\ast}_{\rm{nucleus}}(\mathbf{\Delta}, \mathbf{p}_3, \mathbf{l}')
\end{split}
\end{equation}
where 
\begin{equation}\label{eq:Mnucleus_Fourier}
\begin{split}
&\mathcal{M}_{\rm{nucleus}}(\mathbf{\Delta}, \mathbf{p}_3, \mathbf{l})\mathcal{M}^{\ast}_{\rm{nucleus}}(\mathbf{\Delta}, \mathbf{p}_3, \mathbf{l}') \\
=& \frac{1}{4} g^2\int d^2\mathbf{R} d^2\mathbf{z}_3d^2\mathbf{R}' d^2\mathbf{z}'_3 e^{i(\mathbf{\Delta}+\mathbf{l})\cdot\mathbf{R}} e^{i(\mathbf{p}_3-\mathbf{l}) \cdot\mathbf{z}_3} e^{-i(\mathbf{\Delta}+\mathbf{l}')\cdot\mathbf{R}'} e^{-i(\mathbf{p}_3-\mathbf{l}') \cdot\mathbf{z}'_3}\\
& \qquad \qquad \times A^i_a(\mathbf{R})A^{i'}_b(\mathbf{R}')  \left[U(\mathbf{z}_3)U^{\dagger}(\mathbf{R})-1\right]^{ca}
\left[U(\mathbf{z}'_3)U^{\dagger}(\mathbf{R}')-1\right]^{cb}.
\end{split}
\end{equation}
Eqs.~\eqref{eq:amplitude_squared} and ~\eqref{eq:Mnucleus_Fourier} are the main results in the high momentum limit for diffractive trijet production on the nucleus configuration-by-configuration basis. Averaging over the possible configuration of the nucleus fields, requires finding the expectation value of the above combination of the Wilson lines in the nuclear ensemble, that is  
\begin{equation}\label{eq:WilsonLines_average}
\begin{split}
&\Big\langle  A^i_a(\mathbf{R})A^{i'}_b(\mathbf{R}')  \left[U(\mathbf{z}_3)U^{\dagger}(\mathbf{R})-1\right]^{ca}
\left[U(\mathbf{z}'_3)U^{\dagger}(\mathbf{R}')-1\right]^{cb} \Big\rangle\\
=&\Big\langle  A^i_a(\mathbf{R})A^{i'}_b(\mathbf{R}')  \left[U(\mathbf{z}_3)U^{\dagger}(\mathbf{R})\right]^{ca}
\left[U(\mathbf{z}'_3)U^{\dagger}(\mathbf{R}')\right]^{cb} \Big\rangle\\
&-\Big\langle  A^i_a(\mathbf{R})A^{i'}_b(\mathbf{R}')  
\left[U(\mathbf{z}'_3)U^{\dagger}(\mathbf{R}')\right]^{ab} \Big\rangle -\Big\langle  A^i_a(\mathbf{R})A^{i'}_b(\mathbf{R}')  \left[U(\mathbf{z}_3)U^{\dagger}(\mathbf{R})\right]^{ba}\Big\rangle  
\\
&+\Big\langle  A^i_c(\mathbf{R})A^{i'}_c(\mathbf{R}')\Big\rangle\,.  \\
\end{split}
\end{equation}
The main goal of this study  is the regime where the momentum imbalance  $\mathbf{\Delta}$ forms a very small angle with the momentum of the gluon $\mathbf{p}_3$, since this is the kinematic region where we expect a significant contribution from gluon Bose enhancement. In this regime some terms   in eq.~\eqref{eq:WilsonLines_average} are negligible.  

Specifically the three last terms all are obtained from the first term by putting   $\mathbf{z}_3 =\mathbf{R}$ or $\mathbf{z}'_3 = \mathbf{R}'$ or both. Physically setting  $\mathbf{z}_3 =\mathbf{R}$ means that the gluon and the $q\bar q$ dijet in the amplitude are in the same point in the transverse space. This, in turn, means that the relative momentum between the dijet and the gluon jet is very large. The maximal relative momentum at fixed $|\mathbf{p_3}|$ and $|\mathbf{\Delta}|$ is achieved in the back-to-back configuration of dijet relative to the gluon jet. This term therefore does not contribute to the kinematical region of interest to us.
We verify this conclusion by scrutinizing the result of the explicit calculation for the first term in eq.~\eqref{eq:WilsonLines_average} which we perform below. Substituting   $\mathbf{z}_3 =\mathbf{R}$ or $\mathbf{z}'_3 = \mathbf{R}'$ in this expression leads indeed to a vanishing contribution.

In the following we therefore concentrate on calculating the first term in eq.~\eqref{eq:WilsonLines_average}.

\subsection{Averaging over target}
Our goal now is 
to perform the target averaging of this term. The common practice in CGC calculations is to model the target distribution with the McLerran-Venugopalan (MV) model~\cite{McLerran:1993ka,McLerran:1993ni}. However even within the MV model calculating a correlator like in eq.~\eqref{eq:WilsonLines_average} is  a very complicated endeavor that  we are unable to accomplish without an incommensurate effort. In order to extract physical information from our formulae, and motivated by the structure of a similar calculation in the dilute case,  we are thus led to consider the following factorized approximation
\begin{equation}\label{eq:wilson_line_factorization}
\begin{split}
&\Big\langle  A^i_a(\mathbf{R})A^{i'}_b(\mathbf{R}')  \left[U(\mathbf{z}_3)U^{\dagger}(\mathbf{R})\right]^{ca}
\left[U(\mathbf{z}'_3)U^{\dagger}(\mathbf{R}')\right]^{cb} \Big\rangle\\
\simeq &\Big\langle  A^i_a(\mathbf{R})A^{i'}_b(\mathbf{R}')\Big\rangle \Big\langle   \left[U(\mathbf{z}_3)U^{\dagger}(\mathbf{R})\right]^{ca}
\left[U(\mathbf{z}'_3)U^{\dagger}(\mathbf{R}')\right]^{cb} \Big\rangle\\
&+ \Big\langle  A^i_a(\mathbf{R})  \left[U(\mathbf{z}_3)U^{\dagger}(\mathbf{R})\right]^{ca}\Big\rangle \Big\langle
A^{i'}_b(\mathbf{R}')\left[U(\mathbf{z}'_3)U^{\dagger}(\mathbf{R}')\right]^{cb} \Big\rangle\\
&+\Big\langle  A^i_a(\mathbf{R})\left[U(\mathbf{z}'_3)U^{\dagger}(\mathbf{R}')\right]^{cb}\Big\rangle \Big\langle A^{i'}_b(\mathbf{R}')  \left[U(\mathbf{z}_3)U^{\dagger}(\mathbf{R})\right]^{ca}
 \Big\rangle.\\
\end{split}
\end{equation}
In the limit $|\mathbf{R}-\mathbf{R}'| \sim 1/p_3 \ll 1/Q_s$, the averages in \eqref{eq:wilson_line_factorization} can be explicitly computed in the MV model. For the first term in eq. \eqref{eq:wilson_line_factorization}, one has
\begin{equation}\label{firstterm}
\begin{split}
&\Big\langle  A^i_a(\mathbf{R})A^{i'}_b(\mathbf{R}')\Big\rangle \Big\langle   \left[U(\mathbf{z}_3)U^{\dagger}(\mathbf{R})\right]^{ca}
\left[U(\mathbf{z}'_3)U^{\dagger}(\mathbf{R}')\right]^{cb} \Big\rangle\\
=&G^{ii'}_{WW}(\mathbf{R}, \mathbf{R}') \Big\langle \mathrm{Tr}\left[U(\mathbf{z}_3)U^{\dagger}(\mathbf{R})U(\mathbf{R}')U^{\dagger}(\mathbf{z}'_3)\right]\Big\rangle \\
\simeq &\frac{1}{N_c^2-1}G^{ii'}_{WW}(\mathbf{R}, \mathbf{R}') \Big\langle \mathrm{Tr}\left[U(\mathbf{z}_3)U^{\dagger}(\mathbf{z}'_3)\right]\Big\rangle  \Big\langle \mathrm{Tr}\left[U(\mathbf{R})U^{\dagger}(\mathbf{R}')\right]\Big\rangle\\
=& (N_c^2-1) G^{ii'}_{WW}(\mathbf{R}, \mathbf{R}')D_g(\mathbf{z}_3, \mathbf{z}'_3)D_g(\mathbf{R}, \mathbf{R}') \,.
\end{split}
\end{equation} 
Here we used the factorized approximation for the average of the adjoint quadrupole. This is well justified in our kinematics, since for typical configurations we have the hierarchy of distances $|\mathbf{\Delta}|\sim|\mathbf{R}-\mathbf{R}'|\sim |\mathbf{z}_3- \mathbf{z}'_3|\ll |\mathbf{z}_3-\mathbf{R}|\sim Q_s$. Corrections to this factorization therefore should be of order $Q_s^2/\Delta^2$.

In Eq.~\eqref{firstterm}, $G^{ii'}_{WW}$ is the WW gluon distribution, and $D_g$ is the dipole gluon distribution defined as
\begin{equation}
D_g(\mathbf{x},\mathbf{y})=\frac{1}{N_c^2-1}\langle{\rm Tr}[U^\dagger(\mathbf{x})U(\mathbf{y})]\rangle; \quad G^{ii'}_{WW}(\mathbf{x},\mathbf{y})=\frac{1}{N_c^2-1}\langle A^a_i(\mathbf {x})A^a_{i'}(\mathbf{y})\rangle.
\end{equation}
For both objects, there exist closed analytic expressions in the MV model. Those are given in Appendix B.

The more interesting term is the last term in eq.~\eqref{eq:wilson_line_factorization}. Within the MV model it is computed in Appendix B, see eqs.~\eqref{eq:AUU_one} and~\eqref{eq:AUU_two}: 
\begin{equation}
\begin{split}
&\Big \langle  A^i_a(\mathbf{R})\left[U(\mathbf{z}'_3)U^{\dagger}(\mathbf{R}')\right]^{cb}\Big\rangle\\
=&T^a_{cb} \frac{2 \pi Q_s^2}{N_cg^2}\left[  e^{2 Q_s^2\hat \Gamma_{\mathbf{R}', \mathbf{z}'_3}} \right]  \frac{(-ig)\partial^i_{\mathbf{R}}[\Gamma_{\mathbf{R}, \mathbf{z}'_3} - \Gamma_{\mathbf{R}, \mathbf{R}'}] }{Q_s^2 (\hat \Gamma_{\mathbf{R}, \mathbf{R}'} + \hat  \Gamma_{\mathbf{R}, \mathbf{z}'_3} - \hat  \Gamma_{\mathbf{z}'_3, \mathbf{R}'})}  \left[ e^{ Q_s^2 (\hat \Gamma_{\mathbf{R}, \mathbf{R}'} + \hat \Gamma_{\mathbf{R}, \mathbf{z}'_3} - \hat \Gamma_{\mathbf{z}'_3, \mathbf{R}'})} -1\right] \\
\simeq&T^a_{cb} \frac{4 \pi Q_s^2}{N_cg^2}D_g(\mathbf{R}', \mathbf{z}'_3 ) (-ig)\partial^i_{\mathbf{R}}[L_{\mathbf{R}, \mathbf{z}'_3} -L_{\mathbf{R}, \mathbf{R}'}].
\end{split}
\end{equation}
The approximate equality holds in the high  momentum regime where $|\mathbf{R}-\mathbf{R}'|\sim 1/p_3$ is  small compared to $1/Q_s$. 
The averaging was performed with the MV model, where the correlator of the plus component of the gauge potential is given by
\begin{equation}\label{a+}
\langle A^{+a}(x^-, \mathbf{x})A^{+b}(y^-, \mathbf{y})\rangle 
=\delta^{ab}\delta(x^-- y^-)g^2\mu^2(x^-) L(\mathbf{x}, \mathbf{y})
\end{equation}
with $
L_{\mathbf{x},\mathbf{y}}\equiv L(\mathbf{x}, \mathbf{y}) = \frac{1}{\nabla^4}(\mathbf{x},\mathbf{y}) $. 
The functions $\Gamma$ and $\hat \Gamma$ are defined as
$
\Gamma_{\mathbf{x},\mathbf{y}} = \pi \hat \Gamma_{\mathbf{x},\mathbf{y}} = 2 L(\mathbf{x}, \mathbf{y}) - L(\mathbf{x}, \mathbf{x}) - L(\mathbf{y}, \mathbf{y}) 
$
and we used the conventional definition of the saturation scale 
\begin{equation}Q_s^2 = \frac{N_c}{4\pi} g^4 \int dx^-\mu^2(x^-).\end{equation}

With these expressions we see that the second term in eq. \eqref{eq:wilson_line_factorization} vanishes due to  $T^b_{a a} = 0$:   
\begin{equation}
\Big\langle  A^i_a(\mathbf{R})  \left[U(\mathbf{z}_3)U^{\dagger}(\mathbf{R})\right]^{ca}\Big\rangle  = \Big\langle
A^{i'}_b(\mathbf{R}')\left[U(\mathbf{z}'_3)U^{\dagger}(\mathbf{R}')\right]^{cb} \Big\rangle =0\,.
\end{equation}
We are now ready to compute  eq.~\eqref{eq:Mnucleus_Fourier}
\begin{equation}
\begin{split}
&\Big\langle \mathcal{M}_{\rm{nucleus}}(\mathbf{\Delta}, \mathbf{p}_3, \mathbf{l})\mathcal{M}^{\ast}_{\rm{nucleus}}(\mathbf{\Delta}, \mathbf{p}_3, \mathbf{l}') \Big\rangle  \\
\simeq &\frac{1}{4} g^2\int d^2\mathbf{R} d^2\mathbf{z}_3d^2\mathbf{R}' d^2\mathbf{z}'_3 e^{i(\mathbf{\Delta}+\mathbf{l})\cdot\mathbf{R}} e^{i(\mathbf{p}_3-\mathbf{l}) \cdot\mathbf{z}_3} e^{-i(\mathbf{\Delta}+\mathbf{l}')\cdot\mathbf{R}'} e^{-i(\mathbf{p}_3-\mathbf{l}') \cdot\mathbf{z}'_3}\\
 &\times \Big(\Big\langle  A^i_a(\mathbf{R})A^{i'}_b(\mathbf{R}')\Big\rangle \Big\langle   \left[U(\mathbf{z}_3)U^{\dagger}(\mathbf{R})\right]^{ca}
\left[U(\mathbf{z}'_3)U^{\dagger}(\mathbf{R}')\right]^{cb} \Big\rangle \\
&\qquad +  \Big \langle  A^i_a(\mathbf{R})\left[U(\mathbf{z}'_3)U^{\dagger}(\mathbf{R}')\right]^{cb}\Big\rangle\Big\langle A^{i'}_b(\mathbf{R}')  \left[U(\mathbf{z}_3)U^{\dagger}(\mathbf{R})\right]^{ca}\Big\rangle\Big). \\
\end{split}
\end{equation}
We consider the two terms separately. The first term can be readily written 
\begin{equation}
\begin{split}
&\frac{1}{4} g^2\int d^2\mathbf{R} d^2\mathbf{z}_3d^2\mathbf{R}' d^2\mathbf{z}'_3 e^{i(\mathbf{\Delta}+\mathbf{l})\cdot\mathbf{R}} e^{i(\mathbf{p}_3-\mathbf{l}) \cdot\mathbf{z}_3} e^{-i(\mathbf{\Delta}+\mathbf{l}')\cdot\mathbf{R}'} e^{-i(\mathbf{p}_3-\mathbf{l}') \cdot\mathbf{z}'_3}\\
 &\qquad \times \Big\langle  A^i_a(\mathbf{R})A^{i'}_b(\mathbf{R}')\Big\rangle \Big\langle   \left[U(\mathbf{z}_3)U^{\dagger}(\mathbf{R})\right]^{ca}
\left[U(\mathbf{z}'_3)U^{\dagger}(\mathbf{R}')\right]^{cb} \Big\rangle \\
=& \frac{1}{4} g^2 \Big[(2\pi)^2 \delta(\mathbf{l}-\mathbf{l}')\Big]^2 (N_c^2-1) \int \frac{d^2{\mathbf{k}}}{(2\pi)^2}G^{ii'}_{WW}(-\mathbf{\Delta} -\mathbf{l}-\mathbf{k}) D_g(\mathbf{k})D_g(\mathbf{l}-\mathbf{p}_3)\,.
\end{split}
\end{equation}
For the second term, carrying out the Fourier transformations, one arrives at
\begin{equation}
\begin{split}
 &\frac{1}{4} g^2\int d^2\mathbf{R} d^2\mathbf{z}_3d^2\mathbf{R}' d^2\mathbf{z}'_3 e^{i(\mathbf{\Delta}+\mathbf{l})\cdot\mathbf{R}} e^{i(\mathbf{p}_3-\mathbf{l}) \cdot\mathbf{z}_3} e^{-i(\mathbf{\Delta}+\mathbf{l}')\cdot\mathbf{R}'} e^{-i(\mathbf{p}_3-\mathbf{l}') \cdot\mathbf{z}'_3}\\
 &\qquad \times \Big \langle  A^i_a(\mathbf{R})\left[U(\mathbf{z}'_3)U^{\dagger}(\mathbf{R}')\right]^{cb}\Big\rangle\Big\langle A^{i'}_b(\mathbf{R}')  \left[U(\mathbf{z}_3)U^{\dagger}(\mathbf{R})\right]^{ca}
 \Big\rangle\\
 =&(4 \pi Q_s^2)^2 \frac{N_c}{4g^2}(-ig)^2\int d^2\mathbf{R} d^2\mathbf{z}_3d^2\mathbf{R}' d^2\mathbf{z}'_3 e^{i(\mathbf{\Delta}+\mathbf{l})\cdot\mathbf{R}} e^{i(\mathbf{p}_3-\mathbf{l}) \cdot\mathbf{z}_3} e^{-i(\mathbf{\Delta}+\mathbf{l}')\cdot\mathbf{R}'} e^{-i(\mathbf{p}_3-\mathbf{l}') \cdot\mathbf{z}'_3}\\
 &\qquad \times   D_g(\mathbf{R}'-\mathbf{z}'_3) D_g(\mathbf{R}-\mathbf{z}_3) \partial^i_{\mathbf{R}}[L(\mathbf{R}, \mathbf{z}'_3) - L(\mathbf{R}, \mathbf{R}')]\partial^{i'}_{\mathbf{R}'}[L(\mathbf{R}', \mathbf{z}_3) - L(\mathbf{R}', \mathbf{R})]\\
 =&(4 \pi Q_s^2)^2 \frac{N_c}{4g^2}(-ig)^2 S_{\perp}\int_{\mathbf{q}_1}  \frac{-\mathbf{q}_1^i(-\mathbf{\Delta}-\mathbf{p}_3 +\mathbf{q}_1)^{i'}}{\mathbf{q}_1^4(-\mathbf{\Delta}-\mathbf{p}_3 +\mathbf{q}_1)^4}\Big[ D_g(\mathbf{\Delta} + \mathbf{l} -\mathbf{q}_1) - D_g(\mathbf{l}-\mathbf{p}_3)\Big] \\
 &\qquad \times \Big[ D_g(\mathbf{p}_3-\mathbf{l}'-\mathbf{q}_1) - D_g(\mathbf{p}_3-\mathbf{l}')\Big]\,.
\end{split}
\end{equation}
where the factor of area appears as the limit of the momentum space delta function $S_\perp=(2\pi)^2\delta^2(\mathbf {p}=0)$. The poles at $\mathbf{q}_1 = 0$ and $\mathbf{q}_1 = \mathbf{\Delta} +\mathbf{p}_3$ originate from the Fourier transforms of the relevant factors
\begin{equation}\label{eq:L_difference_prime}
\begin{split}
\partial^i_{\mathbf{R}}[L(\mathbf{R}, \mathbf{z}'_3) - L(\mathbf{R}, \mathbf{R}')] = \int \frac{d^2\mathbf{q}_1}{(2\pi)^2} e^{-i\mathbf{q}_1\cdot \mathbf{R} } \left[e^{+i\mathbf{q}_1\cdot\mathbf{z}'_3} - e^{+i\mathbf{q}_1\cdot\mathbf{R}'}\right] \frac{-i\mathbf{q}_1^i}{\mathbf{q}_1^4}, 
\end{split}
\end{equation}
\begin{equation}\label{eq:L_difference_unprime}
\begin{split}
\partial^{i'}_{\mathbf{R}'}[L(\mathbf{R}', \mathbf{z}_3) - L(\mathbf{R}', \mathbf{R})] =\int \frac{d^2\mathbf{q}_2}{(2\pi)^2} e^{-i\mathbf{q}_2\cdot \mathbf{R}' } \left[e^{+i\mathbf{q}_2\cdot\mathbf{z}_3} - e^{+i\mathbf{q}_2\cdot\mathbf{R}}\right] \frac{-i\mathbf{q}_2^{i'}}{\mathbf{q}_2^4}\,.  
\end{split}
\end{equation}
These poles have  to be regularized at a nonperutrbative scale. In the MV model conventionally this scale is chosen to be  $\Lambda_{\rm QCD}$.  The singularity of the MV model correlator is due to an assumption that the color charges in the transverse plane are uncorrelated. Imposing the condition of color neutrality configuration-by-configuration on the charge density distribution regulates the pole on the spatial scale on which the color neutralization occurs~\footnote{See refs.~\cite{Lam:1999wu,Iancu:2002aq} on the importance of the color neutralization in the CGC.}, which is naturally taken as $\Lambda_{\rm QCD}$. We will follow this practice in this paper, but will comment on the issue more in Section IV.

Putting everything together, the event-averaged diffractive trijet production in the high momentum limit is 
\begin{equation}
\begin{split}
&\langle \left|\mathcal{M}_{\mathrm{diff}} (\mathbf{p}_1, \mathbf{p}_2, \mathbf{p}_3)\right|^2\rangle   \\
=&(2\pi)\frac{4g^2}{N_c}\int \frac{d^2\mathbf{l}}{(2\pi)^2}\frac{d^2\mathbf{l}'}{(2\pi)^2} \frac{\mathbf{l}\cdot\mathbf{l}'}{l^2l^{\prime 2}}\, \partial_{\mathbf{P}_{\perp}^i}\Big[ \Psi^{\gamma^{\ast}\rightarrow q\bar{q}}_{\sigma_1\sigma_2}(p_1^+, p^+_2, \mathbf{P}_{\perp} + \mathbf{l}/2) - \Psi^{\gamma^{\ast}\rightarrow q\bar{q}}_{\sigma_1\sigma_2}(p_1^+, p^+_2, \mathbf{P}_{\perp} - \mathbf{l}/2)\Big]\\
&\times \partial_{\mathbf{P}_{\perp}^{i'}}\Big[ \Psi^{\gamma^{\ast}\rightarrow q\bar{q}}_{\sigma_1\sigma_2}(p_1^+, p^+_2, \mathbf{P}_{\perp} + \mathbf{l}'/2) - \Psi^{\gamma^{\ast}\rightarrow q\bar{q}}_{\sigma_1\sigma_2}(p_1^+, p^+_2, \mathbf{P}_{\perp} - \mathbf{l}'/2)\Big]^{\ast}\\
&\times \Big(S_{\perp}(2\pi)^2 \delta(\mathbf{l}-\mathbf{l}') \frac{(N_c^2-1) g^2}{4} \int \frac{d^2{\mathbf{k}}}{(2\pi)^2}G^{ii'}_{WW}(-\mathbf{\Delta} -\mathbf{l}-\mathbf{k}) D_g(\mathbf{k})
D_g(\mathbf{l}-\mathbf{p}_3)\\
& \qquad - (4\pi Q_s^2)^2 \frac{N_c}{4} S_{\perp}\int \frac{d^2\mathbf{q}_1} {(2\pi)^2} \frac{-\mathbf{q}_1^i(-\mathbf{\Delta}-\mathbf{p}_3 +\mathbf{q}_1)^{i'}}{\mathbf{q}_1^4(-\mathbf{\Delta}-\mathbf{p}_3 +\mathbf{q}_1)^4}\left[ D_g(\mathbf{\Delta} + \mathbf{l} -\mathbf{q}_1) - D_g(\mathbf{l}-\mathbf{p}_3)\right] \\
 &\qquad\qquad  \times \left[ D_g(\mathbf{p}_3-\mathbf{l}'-\mathbf{q}_1) - D_g(\mathbf{p}_3-\mathbf{l}')\right]\Big).
\end{split}
\end{equation}
Note that both terms are of the same order in $g^2$  as $G_{WW}$ has an explicit factor of  $1/g^2$ in its definition, see Appendix B.

\subsection{Summary of analytics}
\label{sec:sa}
Now we summarize our analytic results in the high momentum limit.
The  diffractive trijet production can be written in the factorized form
\begin{equation}\label{eq:factorized_result_large_p}
\begin{split}
&{{ \frac{dN}{d^2\mathbf{p}_1 d^2\mathbf{p}_2 d^2\mathbf{p}_3} \simeq \int \frac{d^2\mathbf{l}}{(2\pi)^2}\frac{d^2\mathbf{l}'}{(2\pi)^2} \sigma^{ii'}(\mathbf{P}_{\perp}, \mathbf{l}, \mathbf{l}') G^{ii'}(\mathbf{\Delta}, \mathbf{p}_3, \mathbf{l}, \mathbf{l}')}}.
\end{split}
\end{equation}
with 
\begin{equation}\label{eq:sigma_part}
\begin{split}
\sigma^{ii'}(\mathbf{P}_{\perp}, \mathbf{l}, \mathbf{l}')
=&(2\pi)\frac{g^4}{N_c} \frac{\mathbf{l}\cdot\mathbf{l}'}{l^2l^{\prime 2}} \partial_{\mathbf{P}_{\perp}^i}\Big[ \Psi^{\gamma^{\ast}\rightarrow q\bar{q}}_{\sigma_1\sigma_2}(p_1^+, p^+_2, \mathbf{P}_{\perp} + \mathbf{l}/2) - \Psi^{\gamma^{\ast}\rightarrow q\bar{q}}_{\sigma_1\sigma_2}(p_1^+, p^+_2, \mathbf{P}_{\perp} - \mathbf{l}/2)\Big]\\
&\times \partial_{\mathbf{P}_{\perp}^{i'}}\Big[ \Psi^{\gamma^{\ast}\rightarrow q\bar{q}}_{\sigma_1\sigma_2}(p_1^+, p^+_2, \mathbf{P}_{\perp} + \mathbf{l}'/2) - \Psi^{\gamma^{\ast}\rightarrow q\bar{q}}_{\sigma_1\sigma_2}(p_1^+, p^+_2, \mathbf{P}_{\perp} - \mathbf{l}'/2)\Big]^{\ast}\\
\end{split}
\end{equation}
and 
\begin{equation}\label{eq:G_part}
\begin{split}
G^{ii'}(\mathbf{\Delta}, \mathbf{p}_3, \mathbf{l}, \mathbf{l}')=&S_{\perp}(2\pi)^2 \delta(\mathbf{l}-\mathbf{l}') \frac{(N_c^2-1)g^2}{4} \int \frac{d^2{\mathbf{k}}}{(2\pi)^2}G^{ii'}_{WW}(-\mathbf{\Delta} -\mathbf{l}-\mathbf{k}) D_g(\mathbf{k}) D_g(\mathbf{l}-\mathbf{p}_3)\\
& - (4\pi Q_s^2)^2 \frac{N_c}{4} S_{\perp}\int \frac{d^2\mathbf{q}_1}{(2\pi)^2}  \frac{-\mathbf{q}_1^i(-\mathbf{\Delta}-\mathbf{p}_3 +\mathbf{q}_1)^{i'}}{\mathbf{q}_1^4(-\mathbf{\Delta}-\mathbf{p}_3 +\mathbf{q}_1)^4}\Big[ D_g(\mathbf{\Delta} + \mathbf{l} -\mathbf{q}_1) - D_g(\mathbf{l}-\mathbf{p}_3)\Big] \\
 &\qquad \times \Big[ D_g(\mathbf{p}_3-\mathbf{l}'-\mathbf{q}_1) - D_g(\mathbf{p}_3-\mathbf{l}')\Big].
\end{split}
\end{equation}
The analytic expressions for the WW  and dipole correlators in coordinate space are given in Appendix \ref{sec:appB}, see eqs.~\eqref{Eq:WWG} and ~\eqref{Eq:Dipole}.

This has to be complemented by the explicit expression for the photon splitting function. 
For the transversely polarized virtual photon, we have~\cite{Beuf:2016wdz} 
\begin{equation}\label{eq:gammaT_WF_1}
\begin{split}
&\Psi^{\gamma^{\ast}\rightarrow q\bar{q}}_{T, \sigma_1\sigma_2}(p_1^+, \mathbf{p}_1; p_2^+, -\mathbf{p}_1) \\
=& -ee_f \delta_{r_\sigma, -\sigma_2} 2\sqrt{z_1z_2} (z_1-z_2+2\lambda \sigma_1) (2\pi) \delta(1-z_1-z_2) \epsilon_{\lambda}^i \mathcal{F}^i(\mathbf{p}_1)\\
\end{split}
\end{equation}
where $-e$ is the electric charge of the electron and $e_f$ represents the quark electric charge of flavor ``$f$'', and we introduced a shorthand  notation 
\begin{equation}
\mathcal{F}^i(\mathbf{p}_1) = \frac{\mathbf{p}_1^i}{\epsilon_f^2 +\mathbf{p}_1^2}
\end{equation}
and  $\epsilon_f^2 = Q^2z_1z_2$. The longitudinal momentum ratios are given by  $z_{1,2} = p_{1,2}^+/q^+$. 

For the longitudinally polarized virtual photon, the wave function is 
\begin{equation}\label{eq:gammaL_WF_1}
\begin{split}
& \Psi^{\gamma^{\ast}\rightarrow q\bar{q}}_{L,\sigma_1\sigma_2}(p_1^+, \mathbf{p}_1; p_2^+, -\mathbf{p}_1) \\
=&-ee_f\delta_{\sigma_1, -\sigma_2} 4(z_1z_2) (2\pi)\delta(1-z_1-z_2) \mathcal{H}(\mathbf{p}_1)\\
\end{split}
\end{equation}
with
\begin{equation}
\mathcal{H}(\mathbf{p}_1) = \frac{\epsilon_f}{\epsilon_f^2+\mathbf{p}_1^2}.
\end{equation}

Squaring eq.~\eqref{eq:gammaT_WF_1} and summing over the spins and polarizations, we can factor out the transverse momentum independent factor
\begin{equation}
\kappa_T(z_1,z_2) = (ee_f)^2 4 (z_1z_2)2[ (z_1-z_2)^2+1] [(2\pi)\delta(1-z_1-z_2)]^2 \,.
\end{equation}
A similar factor for longitudinally polarized photon is
\begin{equation}
\kappa_L(z_1,z_2) = 2(ee_f)^2 16(z_1z_2)^2[(2\pi)\delta(1-z_1-z_2)]^2\,.
\end{equation}

Focusing on the transverse momentum-dependent factors of the dipole wave function, we have for the transverse amplitude
\begin{equation}\label{eq:Mijmn}
\begin{split}
&\mathcal{M}_T^{imn}(\mathbf{l})
\equiv\left(\frac{\mathbf{l}^n}{l^2} \partial_{\mathbf{P}_{\perp}^i} \mathcal{F}^m( \mathbf{P}_{\perp} + \mathbf{l}/2) \right)\\
=&\frac{\mathbf{l}^n}{l^2}\Big[ \delta^{im} \mathcal{H}(\mathbf{P}_{\perp} + \mathbf{l}/2)/\epsilon_f - 2\mathcal{F}^i(\mathbf{P}_{\perp} + \mathbf{l}/2)\mathcal{F}^m (\mathbf{P}_{\perp} + \mathbf{l}/2)\Big]\\
\end{split}
\end{equation}
and for the longitudinal one
\begin{equation}\label{eq:Mijn}
\mathcal{M}_L^{in}(\mathbf{l})
\equiv\left(\frac{\mathbf{l}^n}{l^2} \partial_{\mathbf{P}_{\perp}^i} \mathcal{H}( \mathbf{P}_{\perp} + \mathbf{l}/2) \right)
=\frac{\mathbf{l}^n}{l^2}\Big[- 2 \mathcal{F}^i(\mathbf{P} +\mathbf{l}/2) \mathcal{H}(\mathbf{P}+\mathbf{l}/2)\Big]\,.\\
\end{equation}

Then eq.~\eqref{eq:sigma_part} can be written as
\begin{equation}\label{sigmat}
\sigma_{T}^{ii'}(\mathbf{P}_{\perp}, \mathbf{l}, \mathbf{l}') =2\pi \frac{g^4}{N_c}\kappa_T(z_1, z_2) \Big[ \mathcal{M}_T^{imn}(\mathbf{l}) - \mathcal{M}_T^{imn}(-\mathbf{l})\Big]\Big[ \mathcal{M}_T^{i'mn}(\mathbf{l}') - \mathcal{M}_T^{i'mn}(-\mathbf{l}')\Big]
\end{equation}
for the transverse polarization,
and 
\begin{equation}\label{sigmal}
\sigma_{L}^{ii'}(\mathbf{P}_{\perp}, \mathbf{l}, \mathbf{l}') =2\pi\frac{g^4}{N_c} \kappa_L(z_1, z_2) \Big[ \mathcal{M}_L^{in}(\mathbf{l}) - \mathcal{M}_L^{in}(-\mathbf{l})\Big]\Big[ \mathcal{M}_L^{i'n}(\mathbf{l}') - \mathcal{M}_L^{i'n}(-\mathbf{l}')\Big]
\end{equation}
for the longitudinal polarization. 

Eqs.~\eqref{eq:factorized_result_large_p},~\eqref{eq:G_part},~\eqref{sigmat} and \eqref{sigmal} are the basis for the numerical analysis we present in the next subsection.

\subsection{Numerical results}
For the external momenta $\mathbf{p}_1, \mathbf{p}_2, \mathbf{p}_3$, the trijet production cross section can be expressed in terms of $ \mathbf{P}_{\perp}, \mathbf{\Delta}, \mathbf{p}_3$, 
\begin{equation}
\frac{dN}{d^2\mathbf{P}_{\perp} d^2\mathbf{\Delta} d^2\mathbf{p}_3}=\frac{dN}{d^2\mathbf{p}_1 d^2\mathbf{p}_2 d^2\mathbf{p}_3}\,.
\end{equation}
Since we are interested in the angular correlations between $\mathbf{p}_3$ and $\mathbf{\Delta}$, we integrate over the orientation of the momentum $\mathbf{P}_{\perp}$. We keep the magnitude of $\mathbf{P}_{\perp}$ fixed during this integration. In addition, we also integrate over the direction of  $\mathbf{p}_3$ (denoted by $\beta_3$ below)
\begin{equation}
C(\mathbf{p}_3, \mathbf{\Delta}) = \frac{1}{\mathcal{N}}\int_0^{2\pi} d\beta_3\int d\phi_{P_\perp}  
\frac{dN}{d^2\mathbf{p}_1 d^2\mathbf{p}_2 d^2\mathbf{p}_3}.
\end{equation}
For given $|\mathbf{p}_3|$ and $ |\Delta |$, we plot $C(\mathbf{p}_3, \mathbf{\Delta})$ as a function of the azimuthal angle $\phi$. The normalization factor $1/\mathcal{N}$ representing the normalization over all the angles between $[-\pi/4, \pi/4]$. 

In the numerical computation, strictly speaking one needs to regularize the $1/l^2$ in eq. \eqref{eq:Mijmn}, for which ourpose we use  
\begin{equation}
\frac{1}{l^2} \longrightarrow   \frac{1}{l^2+\Lambda_{\rm QCD}^2}.
\end{equation}
 The factor $\mathbf{l}^n/l^2$ comes from the WW field representing the gluon radiation from the quark and the anti-quark. It turns out that the pole itself contributes only in the terms that do not contain same side correlations, and even in these terms at high momenta the residue of the pole is exponentially small. Thus our results to a very good approximation do not depend on the exact value of the regulator.

We expect that the configuration with small $\mathbf{l}$ (momentum of the emitted gluon) to be most sensitive to Bose enhancement and to be a dominant contributor to the zero-angle  peak. When $\mathbf{l}\rightarrow 0$, the momenta of the gluons inside the nuclear wavefunction probed in this process are $\mathbf{p}_3$ and $\mathbf{\Delta}$, respectively. If the signal is due to Bose enhancement, we expect it to be maximal at\ $\mathbf{p}_3 \approx \mathbf{\Delta}$. 

\begin{figure}
\begin{center}
\includegraphics[width=0.45\linewidth]{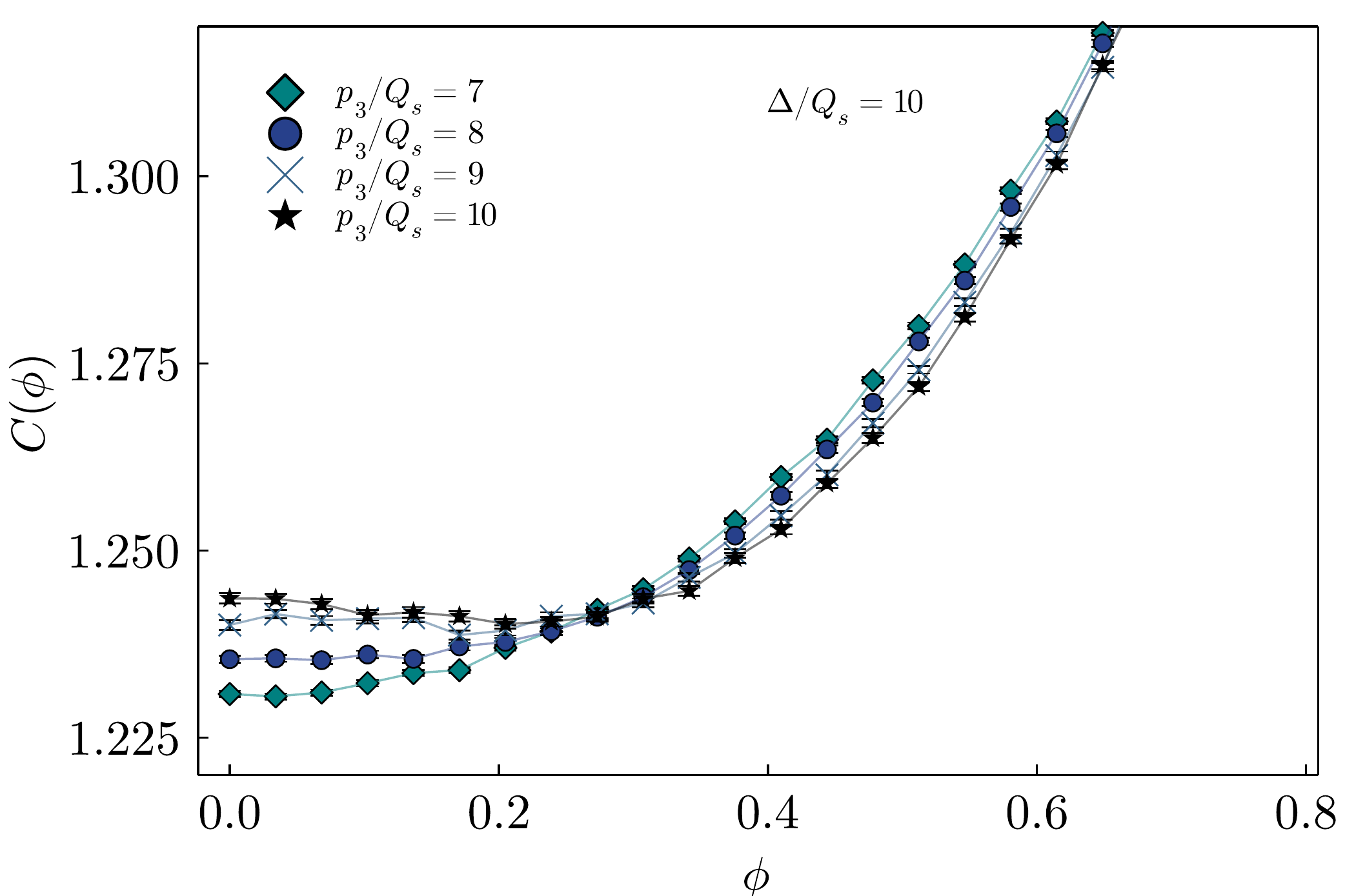}
\includegraphics[width=0.45\linewidth]{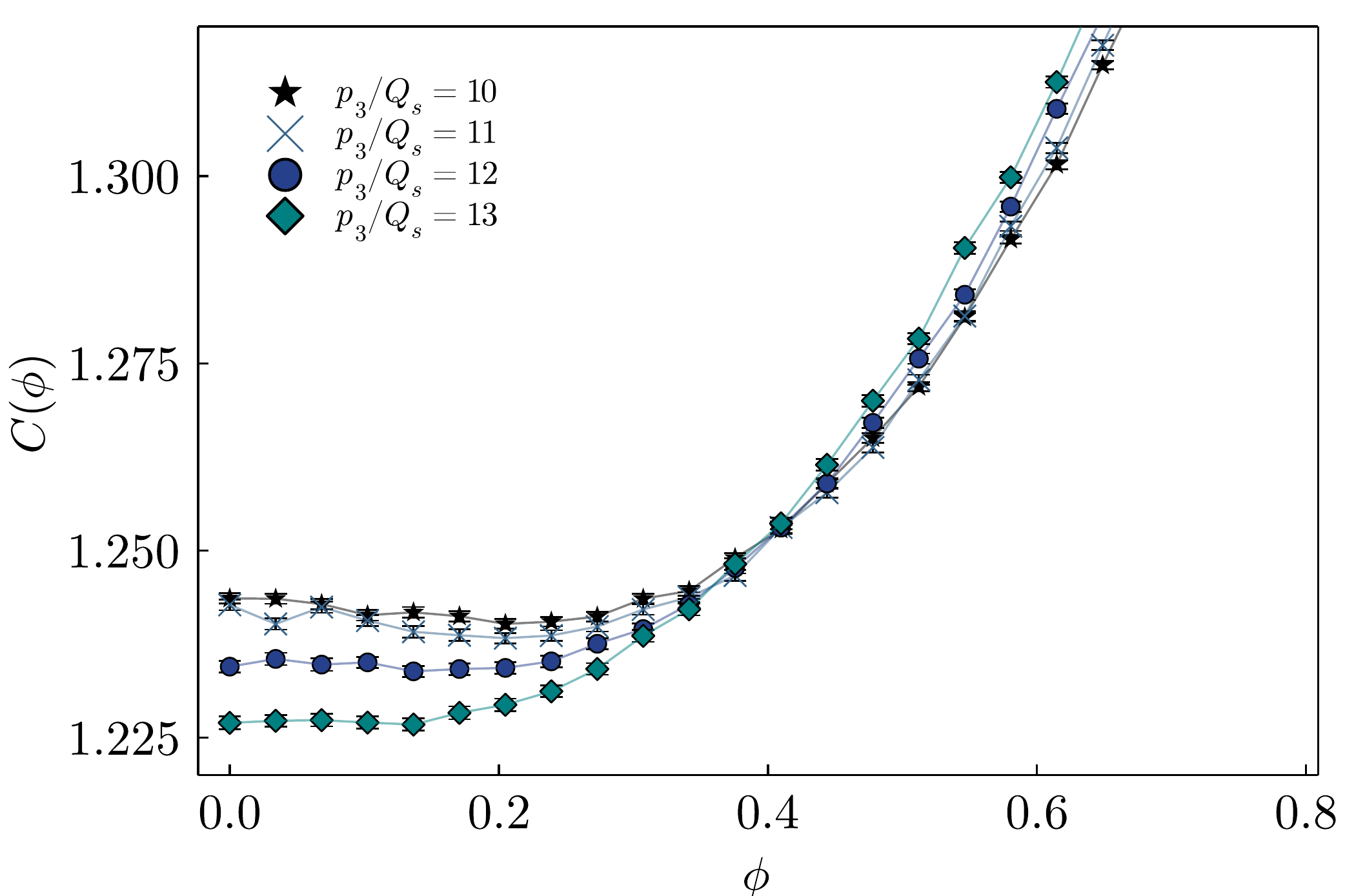}
\caption{The angular correlation function for different values of $p_3$, and $P_\perp/Q_s=15$, $\Delta/Q_s=10$, $z=1/2$, $Q/Q_s=8 $. We consider the transverse polarization of the virtual photon.  The IR scale is chosen to give  $\Lambda_{\rm QCD}/Q_s = 0.2$. }
\label{Fig1}
\end{center}
\end{figure}

\begin{figure}
\begin{center}
\includegraphics[width=0.45\linewidth]{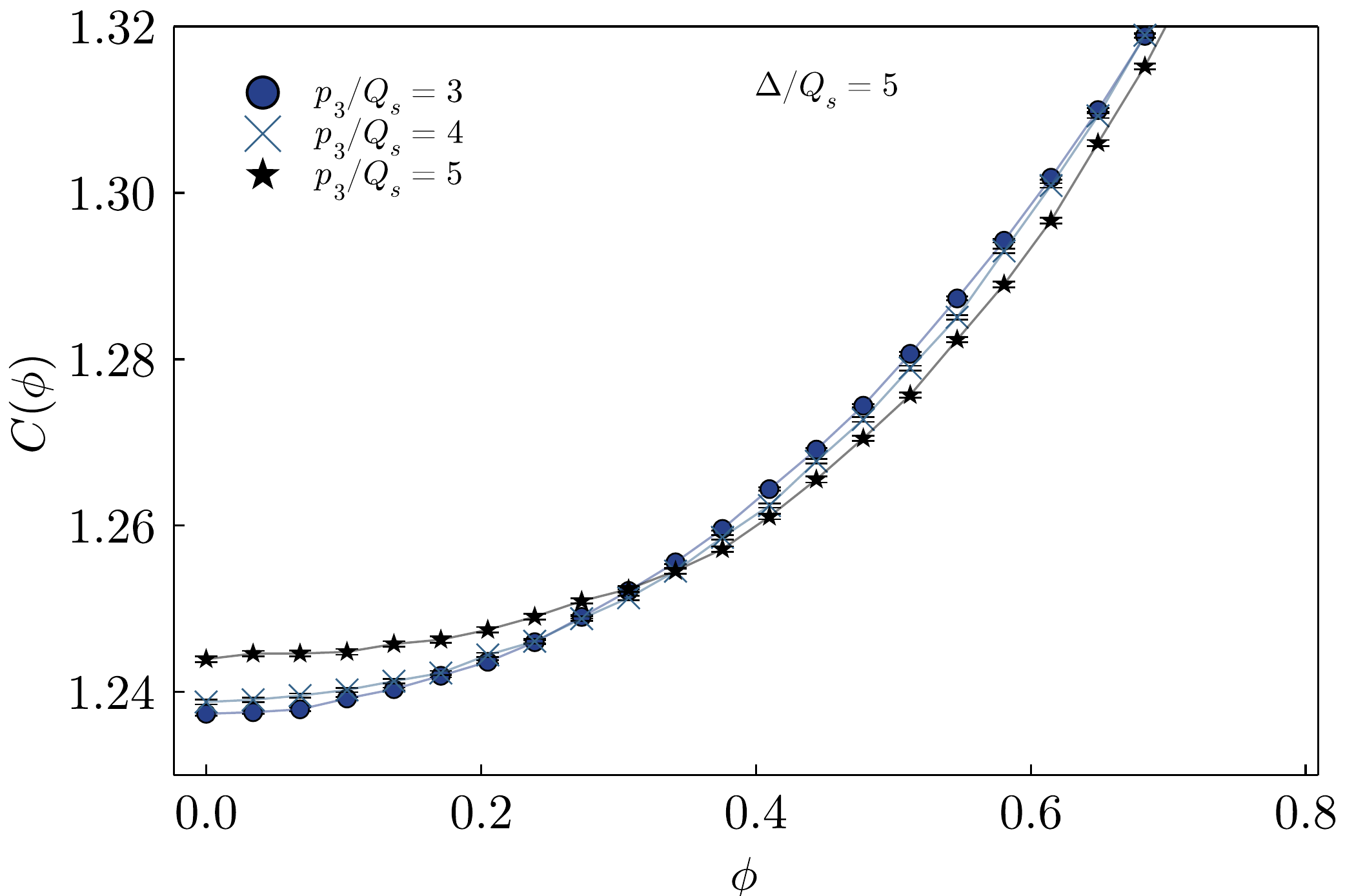}
\includegraphics[width=0.45\linewidth]{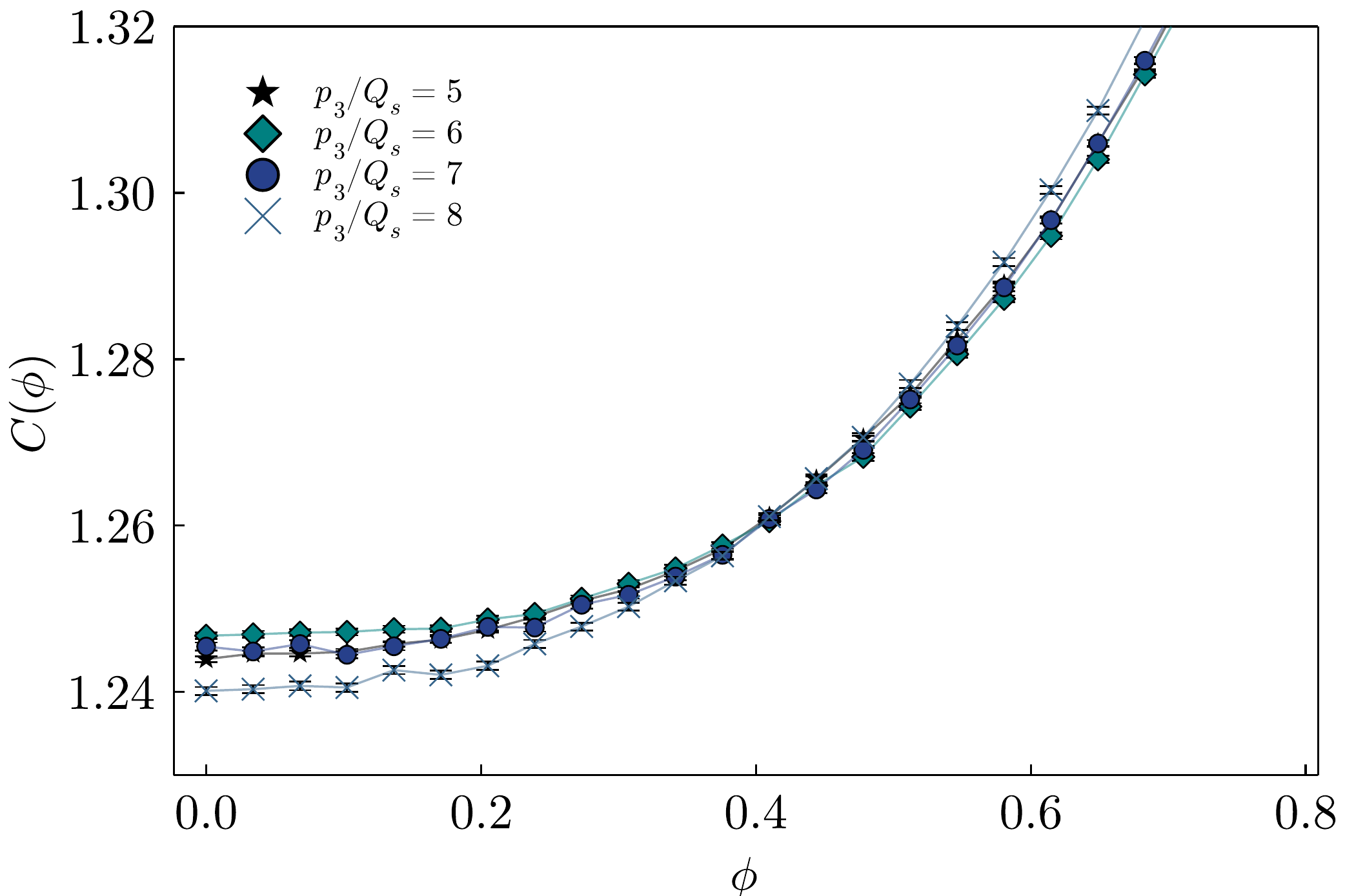}
\caption{The same as fig.~\ref{Fig1}, but for  $P_\perp/Q_s=10$, $\Delta/Q_s=5$, $z=1/2$, $Q/Q_s=8$.}
\label{Fig2}
\end{center}
\end{figure}

Numerically, the calculation is rather complicated as it requires the computation of a six-dimensional integral (integrals with respect to $\mathbf{l}$, $\mathbf{l}'$ and $\mathbf{q}_1$). We performed the integration using Vegas algorithm~\cite{Lepage:1977sw}. The results are presented in figs.~\ref{Fig1} and~\ref{Fig2}.

For very large momenta (fig.~\ref{Fig1}) we observe a clear peak in the correlation function at zero angle. The peak is indeed maximal at $p_3=\Delta$, although it persists in the range $ .8\lesssim p_3/\Delta \lesssim 1.2$. The angular width of the  maximum is about $\Delta\phi\sim .1$, which is consistent with the expectation that "primordial" sharp Bose correlations are somewhat smeared in the scattering by the presence of nonvanishing saturation momentum $Q_s$ leading to a maximum of width $\Delta\phi\sim Q_s/\Delta$.

With decreasing momentum, as shown in fig.~\ref{Fig2}, the peak becomes less pronounced, and at $\Delta/Q_s\sim 5$ it is barely discernible. At this point we cannot tell whether this weakening of the signal is the genuine physical effect, or is due to the fact that we are approaching the applicability limit of the high momentum approximation.

\section{Conclusions}
\label{concl}
In this paper we have studied the diffractive quark-antiquark dijet plus gluon jet production in DIS on a dense target with the aim to see whether this observable is sensitive to Bose correlations of gluons in the hadronic wave function. This is a continuation of our previous work 
\cite{Kovner:2021lty} where we have studied the same observable in the dilute limit. 

We have derived the expression for this observable in terms of the averages of products of Wilson lines. Eq. \eqref{trijetc} is the general expression that is valid in any kinematic regime. Averaging this expression over the target is a very complicated matter and we have not followed this route. Instead we considered the high momentum limit where all momenta are larger than the saturation scale. In this limit we were able to simplify our expression analytically to the point where numerical calculations became feasible. The results of the numerics are shown in figs.~\ref{Fig1} and~\ref{Fig2}.

Our results  confirm the conclusions of \cite{Kovner:2021lty} albeit in a different kinematical regime. We find a noticeable peak at the correlation function at a small angle between $\mathbf{\Delta}$ and $\mathbf{p}_3$ at large transverse momenta. The peak is most pronounced for $|\mathbf{\Delta}|=|\mathbf{p}_3|$ which is consistent with it originating from Bose correlations. The angular width of the peak is of order $\Delta\phi\sim Q_s/\Delta$ which also points to Bose correlations partly smeared by an additional momentum transfer of  around $Q_s$ from the target.

One apparent difference with \cite{Kovner:2021lty} is the dependence on the saturation momentum. While in \cite{Kovner:2021lty} we found that the signal grows with increasing saturation momentum, our present results, figs.~\ref{Fig1} and~\ref{Fig2} 
suggest the opposite trend. The origin of this discrepancy is easily understandable. In \cite{Kovner:2021lty}  we have regulated the infrared pole in the gluon field correlator eq.\eqref{a+} by the saturation momentum $Q_s$. The physical motivation is that in a saturated state, $1/Q_s$ provides the scale of color neutralization and thus has to be the relevant cutoff~\cite{Iancu:2002aq,McLerran:2015sva}. This strong suppression of soft gluons lead to a strong suppression of a back-to-back peak in the trijet production, but did not affect much the same side region in the analysis of \cite{Kovner:2021lty}. Due to this suppression of the back-to-back background, the peak at zero angle became more pronounced. In the current work on the other hand we regulate the same pole with $\Lambda_{QCD}$, as is customarily done in the MV model. This does not suppress the back-to-back peak and thus does not do any favors with the same side signal.

Physically of course one does expect that the color neutralization scale is related to $Q_s$, although this physical effect is outside the realm of the MV model. The strong suppression of the gluon distribution below $Q_s$ is observed for example in the solution of JIMWLK equation \cite{Dumitru:2015gaa}. Thus we expect that this particular effect is taken into account in \cite{Kovner:2021lty} in a more realistic manner than in the present work. Nevertheless, it is interesting to see that even without this suppression, our results show a nice signal at zero angle.

To summarize, the results of the present paper in conjunction with those of \cite{Kovner:2021lty} give a strong indication that the effects of gluon Bose enhancement may be observable in DIS on nuclei.


\section{Acknowledgement}
We are  grateful
to  H.~Duan for stimulating discussions leading to this work.  
We acknowledge the computing resources provided on Henry2, a high-performance computing cluster operated by North Carolina State University, and support by  the U.S. Department of Energy, Office of Science,
Office of Nuclear Physics through the Contract~No.~DE-SC0020081. 
M. Li is supported by the U.S. Department of Energy, Office of Science, Office of Nuclear Physics under Award Number DE-SC0004286.
A.K. is supported by the NSF Nuclear Theory grant 2208387. This material is based
upon work supported by the U.S. Department of Energy, Office of Science, Office of Nuclear
Physics through the Saturated Glue (SURGE) Topical Collaboration.

\appendix
\section{Dilute target limit}
In the limit of the dilute target, the diffractive dijet plus gluon jet production should be expanded to the first nontrivial order in powers of target color charge density (or target color field). This involves an expansion of the Wilson lines. 

Consider the diffractive production amplitude Eq.~\eqref{eq:diffractive_amplitude}. First note that both the quark and the antiquark Wilson lines have to be expanded at least to order-$g$, otherwise, the trace over fundamental Wilson vanishes. 
\begin{equation}
\begin{split}
 &\mathcal{U}^{\dagger ac_3}(\mathbf{z}_3) \mathrm{Tr}\left[\mathcal{S}(\mathbf{z}_2) t^a \mathcal{S}^{\dagger}(\mathbf{z}_1)\right]-\mathrm{Tr}\left[\mathcal{S}(\mathbf{z}_2)\mathcal{S}^{\dagger}(\mathbf{z}_1)t^{c_3}\right]\\
 \approx&-\frac{1}{2} g^2 T^{c_3}_{ab}\Big[ \alpha_T^a(\mathbf{z}_3)\alpha_T^b(\mathbf{z}_2)-\alpha_T^{a}(\mathbf{z}_3)\alpha_T^b(\mathbf{z}_1) - \alpha_T^a(\mathbf{z}_1)\alpha_T^b(\mathbf{z}_2)\Big]+ \mathcal{O}(g^3)\\
=&-\frac{1}{2} g^2 T^{c_3}_{ab}\int \frac{d^2\mathbf{k}}{(2\pi)^2}\frac{d^2\mathbf{k}'}{(2\pi)^2}  \Big[ e^{-i\mathbf{k}\cdot\mathbf{z}_3-i\mathbf{k}'\cdot\mathbf{z}_2} -e^{-i\mathbf{k}\cdot\mathbf{z}_3 -i\mathbf{k}'\cdot\mathbf{z}_1 } - e^{-i\mathbf{k}\cdot\mathbf{z}_1 -i\mathbf{k}'\cdot\mathbf{z}_2}\Big]\alpha_T^a(\mathbf{k})\alpha_T^b(\mathbf{k}')
\end{split}
\end{equation}
Proceeding to the other factors in Eq.~\eqref{eq:diffractive_amplitude}, we have 
\begin{equation}
\Big[ \partial_j \phi(\mathbf{z}_3-\mathbf{z}_2) - \partial_j\phi(\mathbf{z}_3-\mathbf{z}_1)\Big]=\int \frac{d^2\mathbf{l}}{(2\pi)^2}\frac{i\mathbf{l}_j}{l_{\perp}^2 } \Big[ e^{-i\mathbf{l}\cdot(\mathbf{z}_3-\mathbf{z}_2)} -  e^{-i\mathbf{l}\cdot(\mathbf{z}_3-\mathbf{z}_1)}\Big]
\end{equation}
and 
\begin{equation}\label{eq:fourier_three}
\Psi^{\gamma^{\ast}\rightarrow q\bar{q}}_{\sigma_1\sigma_2}(p_1^+, \mathbf{z}_1; p_2^+, \mathbf{z}_2)
=\int \frac{d^2\mathbf{k}_1}{(2\pi)^2} e^{-i\mathbf{k}_1\cdot(\mathbf{z}_1-\mathbf{z}_2)} \Psi^{\gamma^{\ast}\rightarrow q\bar{q}}_{\sigma_1\sigma_2}(p_1^+, \mathbf{k}_1; p_2^+,- \mathbf{k}_1).
\end{equation}
Using these expressions, 
the trijet production amplitude in momentum space becomes
\begin{equation}\label{eq:amplitude_dilute_f1}
\begin{split}
&\mathcal{M}(\mathbf{p}_1, \mathbf{p}_2, \mathbf{p}_3)=\int d^2\mathbf{z}_1 d^2\mathbf{z}_2 d^2\mathbf{z}_3 e^{i\mathbf{p}_1\cdot\mathbf{z}_1}  e^{i\mathbf{p}_2\cdot\mathbf{z}_2}  e^{i\mathbf{p}_3\cdot\mathbf{z}_3} \mathcal{M}(\mathbf{z}_1, \mathbf{z}_2, \mathbf{z}_3)\\
=&i\frac{g^3}{\sqrt{N_c}}\sqrt{2\pi} T^{c_3}_{ab} \int \frac{d^2\mathbf{k}'}{(2\pi)^2}\alpha^a_T(\mathbf{p}-\mathbf{k}')\alpha_T^b(\mathbf{k}')\Big\{ \frac{i(\mathbf{k}' - \mathbf{\Delta})_j}{|\mathbf{k}'-\mathbf{\Delta}|^2}\Big[ \Psi^{\gamma^{\ast}\rightarrow q\bar{q}}_{\sigma_1\sigma_2} (p_1^+, \mathbf{p}_1; p_2^+, -\mathbf{p}_1)\\
&-\Psi^{\gamma^{\ast}\rightarrow q\bar{q}}_{\sigma_1\sigma_2} (p_1^+, \mathbf{p}_1-\mathbf{k}'; p_2^+, -\mathbf{p}_1+\mathbf{k}')-\Psi^{\gamma^{\ast}\rightarrow q\bar{q}}_{\sigma_1\sigma_2} (p_1^+, -\mathbf{p}_2+\mathbf{k}'; p_2^+, \mathbf{p}_2-\mathbf{k}')+\Psi^{\gamma^{\ast}\rightarrow q\bar{q}}_{\sigma_1\sigma_2} (p_1^+, -\mathbf{p}_2; p_2^+, \mathbf{p}_2)\Big]\\
&+\frac{i\mathbf{p}_{3,j}}{p_3^2}\Big[ \Psi^{\gamma^{\ast}\rightarrow q\bar{q}}_{\sigma_1\sigma_2} (p_1^+, \mathbf{p}_1-\mathbf{k}'; p_2^+, -\mathbf{p}_1+\mathbf{k}')+\Psi^{\gamma^{\ast}\rightarrow q\bar{q}}_{\sigma_1\sigma_2} (p_1^+, -\mathbf{p}_2+\mathbf{k}'; p_2^+, \mathbf{p}_2-\mathbf{k}')\Big]\Big\}
\end{split}
\end{equation}
Next, we introduce the Lipatov vertex for the single gluon production  (note that $\mathbf{p}_3$ is the momentum of the produced gluon)
\begin{equation}
L_j(-(\mathbf{p}_1+\mathbf{p}_2+\mathbf{k}), \mathbf{p}_3) = -\left(\frac{(\mathbf{p}_1+\mathbf{p}_2+\mathbf{k})_j}{|\mathbf{p}_1+\mathbf{p}_2+\mathbf{k}|^2} +\frac{\mathbf{p}_{3,j}}{p_3^2}\right)  = \left(\frac{\mathbf{k}'_j}{|\mathbf{k}'|^2} -\frac{\mathbf{p}_{3,j}}{|\mathbf{p}_3|^2}\right)
\end{equation}
where $\mathbf{k}' \equiv -(\mathbf{p}_1+\mathbf{p}_2+\mathbf{k})$.

\begin{figure}[t]
\centering 
\includegraphics[width = 0.9\textwidth]{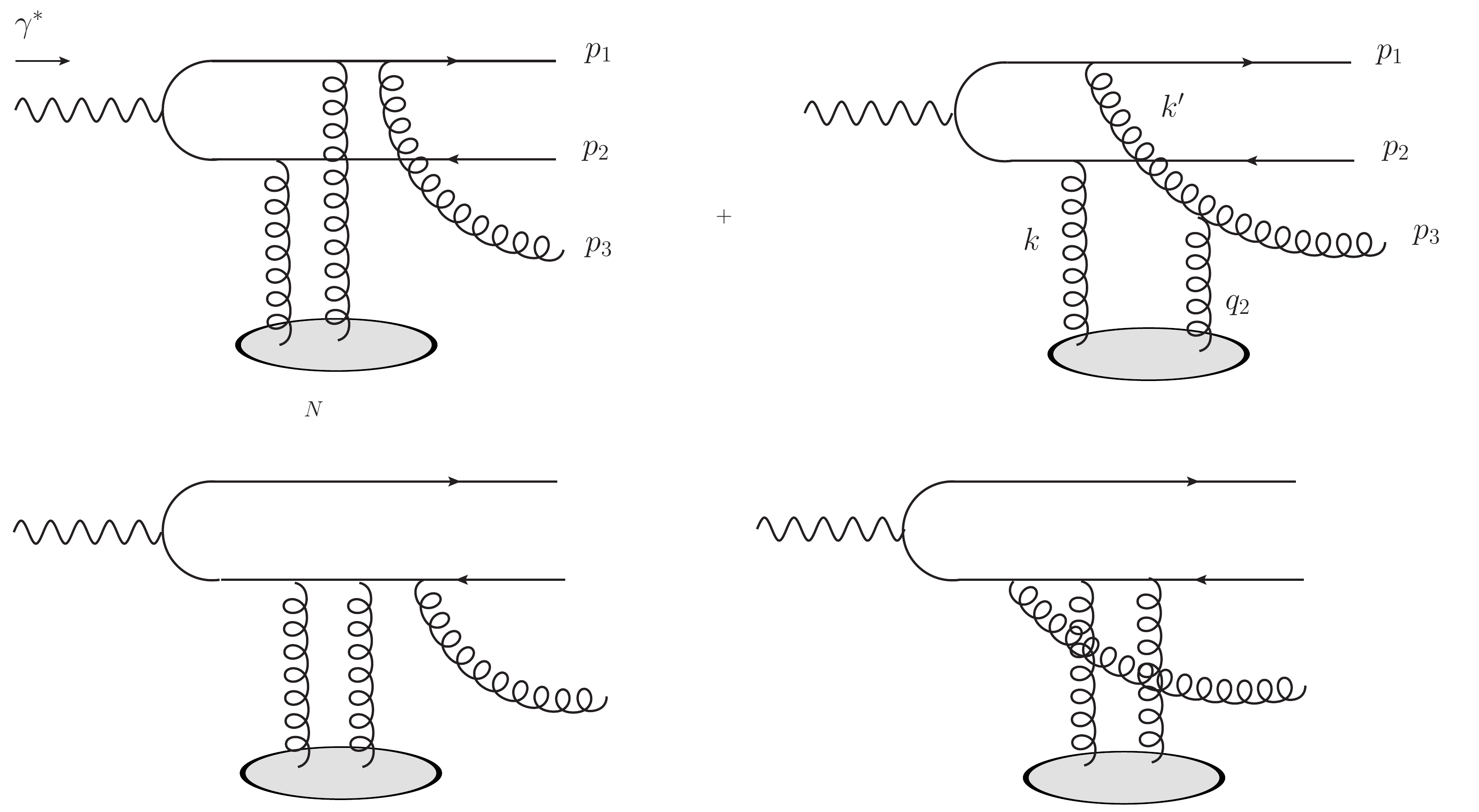}
\caption{Schematic diagram illustrating the diffractive trijet production with the gluon jet in the electron-going direction. In the dilute limit, two gluon exchanges is the lowest order. }
\label{fig:trijet_lipatov_vertex}
\end{figure} 

\begin{figure}[!t]
    \centering
    \includegraphics[width=0.9\textwidth]{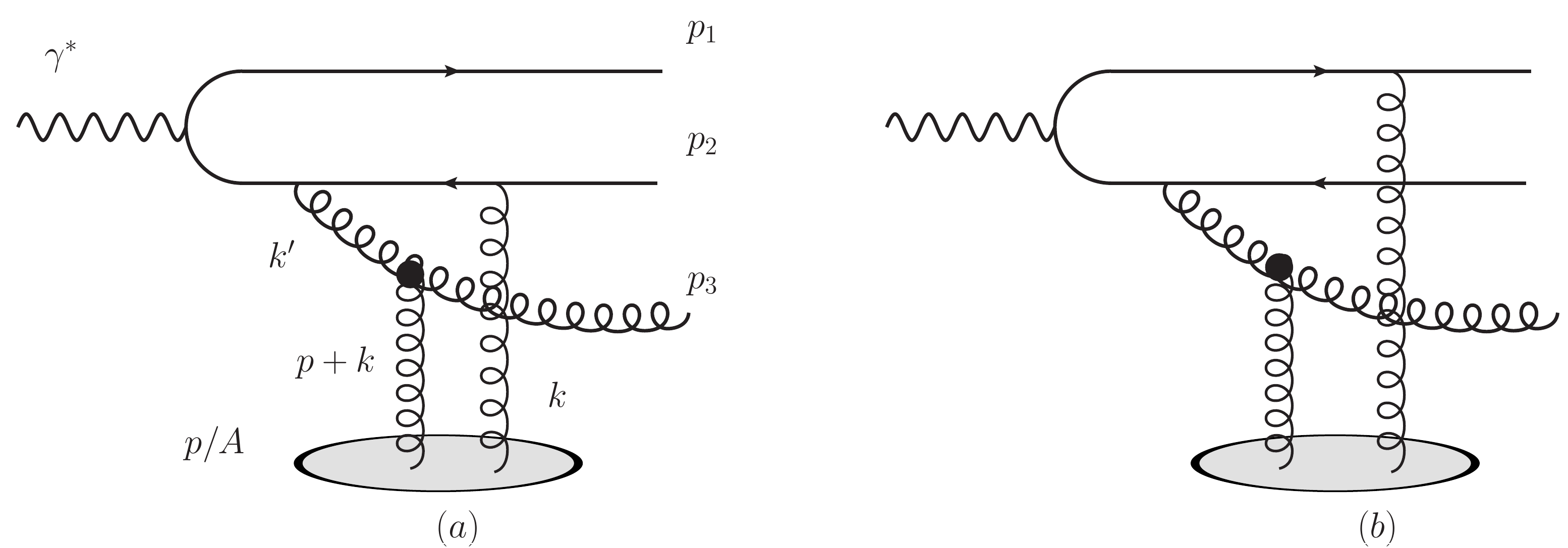}
    \caption{
		Schematic diagram showing the trijet production in $\gamma^{\ast}N$ collisions. The black solid circle represents the effective Lipatov vertex. There are two more similar diagrams related by switching the roles of quark and antiquark.  $\mathbf{k}' = -(\mathbf{p}_1+\mathbf{p}_2+\mathbf{k})$ and $\mathbf{p} =\mathbf{p}_1+\mathbf{p}_2+\mathbf{p}_3$. }  
		\label{fig:diag_feynman}
\end{figure}

It is convenient to introcude a short-hand notation 
\begin{equation}
  \label{eq:dipole_wf_diff}
\begin{split}
\psi^D_{\sigma_1\sigma_2}(\mathbf{p}_1, \mathbf{p}_2, \mathbf{k}) =&\Big[ \Psi^{\gamma^{\ast}\rightarrow q\bar{q}}_{\sigma_1\sigma_2} (p_1^+, \mathbf{p}_1; p_2^+, -\mathbf{p}_1)-\Psi^{\gamma^{\ast}\rightarrow q\bar{q}}_{\sigma_1\sigma_2} (p_1^+, \mathbf{p}_1+\mathbf{k}; p_2^+, -\mathbf{p}_1-\mathbf{k})\\
&-\Psi^{\gamma^{\ast}\rightarrow q\bar{q}}_{\sigma_1\sigma_2} (p_1^+, -\mathbf{p}_2-\mathbf{k}; p_2^+, \mathbf{p}_2+\mathbf{k})+\Psi^{\gamma^{\ast}\rightarrow q\bar{q}}_{\sigma_1\sigma_2} (p_1^+, -\mathbf{p}_2; p_2^+, \mathbf{p}_2)\Big].\\ 
\end{split}
\end{equation}
Then the amplitude in Eq.~\eqref{eq:amplitude_dilute_f1} can be written as 
\begin{equation}
  \label{eq:amplitude_final_res}
\begin{split}
&\mathcal{M}(\mathbf{p}_1, \mathbf{p}_2, \mathbf{p}_3)\\
=&\frac{g^3}{\sqrt{N_c}}\sqrt{2\pi} T^{c_3}_{ab} \int \frac{d^2\mathbf{k}}{(2\pi)^2}\alpha^a_T(\mathbf{p}+\mathbf{k})\alpha_T^b(-\mathbf{k}) L_j(-(\mathbf{p}_1+\mathbf{p}_2+\mathbf{k}), \mathbf{p}_3) \psi_{\sigma_1\sigma_2}^D(\mathbf{p}_1, \mathbf{p}_2, \mathbf{k})\\
=&g^3\sqrt{2\pi} T^{c_3}_{ab} \int \frac{d^2\mathbf{k}}{(2\pi)^2}\rho^a_T(\mathbf{p}+\mathbf{k})\rho_T^b(-\mathbf{k}) \frac{L_j(-(\mathbf{p}_1+\mathbf{p}_2+\mathbf{k}), \mathbf{p}_3)}{k^2_{\perp} |\mathbf{p}+\mathbf{k}|^2} \psi_{\sigma_1\sigma_2}^D(\mathbf{p}_1, \mathbf{p}_2, \mathbf{k})\\
\end{split}
\end{equation}
This expression is easy to understand. There are four terms in eq. \eqref{eq:dipole_wf_diff}; thus \eqref{eq:amplitude_final_res} can be illustrated by four diagrams, see fig.~\ref{fig:trijet_lipatov_vertex}.  Fig.~\ref{fig:diag_feynman} shows the equivalent diagrams using the Lipatov vertex.  It is important to note that transverse momentum conservation imposes $\mathbf{k}' = -(\mathbf{p}_1+\mathbf{p}_2+\mathbf{k})$ and there is only one independent momentum that is integrated over. 

The trijet production amplitude squared in the dilute limit becomes   
\begin{equation}
\begin{split}
&|\mathcal{M}(\mathbf{p}_1, \mathbf{p}_2, \mathbf{p}_3)|^2\\
 =&- (2\pi)\frac{g^6}{N_c} T^{c_3}_{ab}T^{c_3}_{cd} \int\frac{d^2\mathbf{k}}{(2\pi)^2}\frac{d^2\mathbf{l}}{(2\pi)^2} \rho_T^a(\mathbf{p}+\mathbf{k}) \rho_T^b(-\mathbf{k}) \rho_T^c(-\mathbf{p}-\mathbf{l})\rho_T^d(\mathbf{l}) \frac{1}{k_{\perp}^2l_{\perp}^2} \\
&\times \frac{1}{|\mathbf{p}+\mathbf{k}|^2|\mathbf{p} + \mathbf{l}|^2}L_j(-(\mathbf{p}_1+\mathbf{p}_2+\mathbf{k}), \mathbf{p}_3)L_j(-(\mathbf{p}_1+\mathbf{p}_2+\mathbf{l}), \mathbf{p}_3) \psi_{\sigma_1\sigma_2}^D(\mathbf{p}_1, \mathbf{p}_2, \mathbf{k}) \psi_{\sigma_1\sigma_2}^{D \ast}(\mathbf{p}_1, \mathbf{p}_2, \mathbf{l}).
\end{split}
\end{equation}
This is the same as the result as we obtained in ref.~\cite{Kovner:2021lty} (apart from the overall factor $1/N_c$ which was missed in \cite{Kovner:2021lty}).

\section{Wilson line correlators in the MV model}
\label{sec:appB}

In this appendix, we will consider a few Wilson line correlators required for our calculations. We will use the MV model, which approximates the correlations of the 
covariant gauge color charge densities by a Gaussian distribution with 
\begin{equation}
\langle \tilde{\rho}^a(x^-, \mathbf{x}_{\perp}) \tilde{\rho}^b(y^-, \mathbf{y}_{\perp})\rangle = \delta^{ab}\delta(x^--y^-)\delta(\mathbf{x}_{\perp}-\mathbf{y}_{\perp}) g^2\mu^2(x^-)\,.
\end{equation}
Here we will consider $\mu^2(x^-)$ to be independent of the transverse coordinates. An  extension to an arbitrary  transverse coordinate dependent $\mu^2(x^-,\mathbf{x}_\perp)$ is in principle straightforward. 

In the covariant gauge, the only nonvanishing component of the gluon field is 
\begin{equation}
 \tilde{A}^{+,a}(x^-, \mathbf{x}_{\perp}) = \int d^2\mathbf{z}_{\perp} G_0(\mathbf{x}_{\perp}-\mathbf{z}_{\perp})\tilde{\rho}^a(x^-, \mathbf{z}_{\perp})\,,
\end{equation}
where the Green function is $G_0(\mathbf{x}_{\perp}-\mathbf{z}_{\perp}) = -1/\nabla^2$.  As a consequence, one obtains
\begin{equation}
\langle \tilde{A}^{+,a}(x^-, \mathbf{x}_{\perp}) \tilde{A}^{+,a}(y^-, \mathbf{y}_{\perp})    \rangle 
=\delta^{ab}\delta(x^-- y^-)g^2\mu^2(x^-) L(\mathbf{x}_{\perp}, \mathbf{y}_{\perp})
\end{equation}
Here 
\begin{equation}
L(\mathbf{x}_{\perp}, \mathbf{y}_{\perp}) = \int d^2\mathbf{z}_{\perp} G_0(\mathbf{x}_{\perp}-\mathbf{z}_{\perp})G_0(\mathbf{y}_{\perp}-\mathbf{z}_{\perp}).
\end{equation}
In the light-cone gauge $A^+ =0$, it is the transverse components of the vector potential that do not vanish 
\begin{equation}
  \label{Eq:Ai}
A^i_a(x^-, \mathbf{x}_{\perp}) = \int_{-\infty}^{x^-} dz^- \partial^+ A^i_a(z^-, \mathbf{x}_{\perp})=\int_{-\infty}^{x^-} dz^- F^{+i}_a(z^-, \mathbf{x}_{\perp})
\end{equation}
where 
where the boundary condition $A^i(x^-=-\infty, \mathbf{x}_{\perp}) =0$ is consistent with fixing the residual gauge at minus infinity. The light-cone gauge field strength tensor is related to the covariant gauge field strength tensor (labeled by tilde) through a gauge transformation
\begin{equation}
  \label{Eq:GaugeTr}
F^{+i}_a(x^-, \mathbf{x}_{\perp}) = U_{ac}^{\dagger}(x^-, \mathbf{x}_{\perp})\widetilde{F}^{+i}_c(x^-, \mathbf{x}_{\perp}) 
\end{equation}
where  the adjoint representation Wilson line is given by 
\begin{equation}
U_{ac}(x^-, \mathbf{x}_{\perp}) = \mathcal{P} \mathrm{exp}\left\{-ig\int_{-\infty}^{x^-} dz^- \tilde{A}^{+,a}(z^-, \mathbf{x}_{\perp})T^a\right\}\,.
\end{equation}

\subsection{The field correlator $\langle A_iA_j\rangle $}
For reference, we reproduce the WW correlator $\langle A_iA_j\rangle$ in this section. 
Using eqs.~\eqref{Eq:Ai} and~\eqref{Eq:GaugeTr}, the gauge field correlator can be expressed in terms of quantities in the covariant gauge (we follow ref.~\cite{Jalilian-Marian:1996mkd})
\begin{equation}
\begin{split}
& \Big\langle A^i_a(x^-, \mathbf{x}_{\perp}) A^j_b(y^-, \mathbf{y}_{\perp}) \Big\rangle\\
  = &\int_{-\infty}^{x^-}dz^- \int_{-\infty}^{y^-} dw^-\Big \langle U^{\dagger}_{ac}(z^-, \mathbf{x}_{\perp})\partial^i \tilde{A}^+_c(z^-, \mathbf{x}_{\perp}) U^{\dagger}_{bd}(w^-, \mathbf{y}_{\perp})\partial^j \tilde{A}^+_d(w^-, \mathbf{y}_{\perp}) \Big\rangle\\
  =&\int_{-\infty}^{x^-}dz^- \int_{-\infty}^{y^-} dw^- \langle U_{ca}(z^-, \mathbf{x}_{\perp})U_{db}(w^-, \mathbf{y}_{\perp})\rangle \partial^i_{\mathbf{x}}\partial^j_{\mathbf{y}} \langle \tilde{A}_c^+(z^-, \mathbf{x}_{\perp}) \tilde{A}_d^+(w^-, \mathbf{y}_{\perp})\rangle \\
  =&\frac{\delta^{ab}\partial^i_{\mathbf{x}}\partial^j_{\mathbf{y}} L(\mathbf{x}_{\perp}, \mathbf{y}_{\perp})}{\frac{1}{2} N_c g^2 \Gamma(\mathbf{x}, \mathbf{y})} \Big[ D_g(\mathbf{x}-\mathbf{y}, \mathrm{min}\{x^-, y^-\}) -1\Big]
  \end{split} 
\end{equation}
Here 
$ \Gamma(\mathbf{x}_{\perp}, \mathbf{y}_{\perp}) = 2 L(\mathbf{x}_{\perp}, \mathbf{y}_{\perp}) - L(\mathbf{x}_{\perp}, \mathbf{x}_{\perp}) - L(\mathbf{y}_{\perp}, \mathbf{y}_{\perp})$ 
and $D$ is 
the dipole correlator for adjoint representation
\begin{equation}
  \label{Eq:Dipole}
D_g(\mathbf{x}-\mathbf{y}, z^-) = \mathrm{exp}\left\{\frac{1}{2}N_c g^4 \bar{\mu}^2(z^-) \Gamma(\mathbf{x}, \mathbf{y})\right\}.
\end{equation}
with $\bar{\mu}^2(z^-) = \int_{-\infty}^{z^-}dv^- \mu^2(v^-)$.  For $\mathrm{min}\{x^-, y^-\}>0$, one obtains
\begin{equation}
\Big\langle A^i_a(\mathbf{x}_{\perp}) A^j_b( \mathbf{y}_{\perp}) \Big\rangle =  \delta^{ab}G_{WW}^{ij}(\mathbf{x}_{\perp},\mathbf{y}_{\perp}) =\delta^{ab}\frac{\partial^i_{\mathbf{r}}\partial^j_{\mathbf{r}} L(\mathbf{r})}{\frac{1}{2} N_c g^2 \Gamma(\mathbf{r})}  \left[e^{\frac{1}{2}N_c g^4\bar{\mu}^2 \Gamma(\mathbf{r})} -1\right] 
\end{equation}
with
\begin{equation}
\partial^i_{\mathbf{r}}\partial^j_{\mathbf{r}} L(\mathbf{r}) = - \int \frac{d^2\mathbf{p}}{(2\pi)^2} e^{-i\mathbf{p}\cdot\mathbf{r}} \frac{\mathbf{p}^i\mathbf{p}^j}{p_{\perp}^4}. 
\end{equation}
Note that this expression requires regularization in the infrared. 
Similarly, 
\begin{equation}
\begin{split}
\Gamma(\mathbf{r}) = 2L(\mathbf{r})-2L(0) = &2\int \frac{d^2\mathbf{p}}{(2\pi)^2} [e^{-i\mathbf{p}\cdot\mathbf{r}}-1] \frac{1}{p_{\perp}^4}\\
\simeq & -\frac{1}{4\pi} r^2 \ln\left[ \frac{1}{ \Lambda_{\rm QCD} r}  \right] \,.
\end{split}
\end{equation}
Here the IR scale $\Lambda_{\rm QCD}$ enters by replacing the propagator $1/p^2$ with $1/(p^2+\Lambda_{\rm QCD}^2)$.   For convenience, we also define $\hat \Gamma(r) = \pi  \Gamma(r)$.  

For the purpose of the numerics, the momentum space expression is required for the field correlator. To derive its semi-analytic form, 
we start with the coordinate space expression 
\begin{equation}
  \label{Eq:WWG}
\begin{split}
G_{WW}^{ij}(r) = &\frac{-\partial^i_{\mathbf{r}} \partial^j_{\mathbf{r}} \Gamma(r)}{N_cg^2\Gamma(r)}\left[ e^{2 Q_s^2\hat \Gamma(r)}-1\right]\\
=&\frac{-2}{N_cg^2}\frac{1}{\Gamma(r)}\left( \delta^{ij} \frac{d}{dr^2} \Gamma(r) + 2\mathbf{r}^i\mathbf{r}^j \left(\frac{d}{dr^2}\right)^2 \Gamma(r)\right) \left[ e^{2 Q_s^2\hat \Gamma(r)}-1\right]\,.
\end{split}
\end{equation}
Here we used the conventional definition of the saturation scale $Q_s^2 = \frac{1}{4\pi} N_c g^4 \bar{\mu}^2$. 
Note that $\Gamma(r)$ can be written as a function of $r^2$ and the derivatives are with respect to $r^2$. Details of the computation are given in ref.~\cite{Dumitru:2016jku}. 

The trace part of the tensor $G_{WW}^{ij}$ is given by  
\begin{equation}
G^{(1)}(p) = \delta^{ij} \int d^2\mathbf{r} e^{i\mathbf{p}\cdot\mathbf{r}} G_{WW}^{ij} (r) 
=\frac{-8\pi }{N_cg^2} \int dr r J_0(pr)\frac{1}{r^2} \left[ e^{2 Q_s^2\hat \Gamma(r)}-1\right]
\end{equation}
while 
the traceless part 
\begin{equation}
\begin{split}
h^{(1)}(p) =& -\left(\delta^{ij} - \frac{2\mathbf{p}^i \mathbf{p}^j}{p^2}\right) \int d^2\mathbf{r} e^{i\mathbf{p}\cdot\mathbf{r}} G_{WW}^{ij} (r) \\
=&\frac{4}{N_cg^2} \int d^2\mathbf{r} e^{i\mathbf{p}\cdot\mathbf{r}} (1-2\cos^2{\theta} )r^2\frac{1}{\Gamma(r)}\left((\frac{d}{dr^2})^2 \Gamma(r)\right) \left[ e^{2 Q_s^2\hat \Gamma(r)}-1\right]\\
=&\frac{8\pi }{N_cg^2} \int dr rJ_2(pr) r^2\frac{1}{\Gamma(r)}\left((\frac{d}{dr^2})^2 \Gamma(r)\right) \left[ e^{2 Q_s^2\hat \Gamma(r)}-1\right]\\
\simeq &\frac{8\pi }{N_cg^2} \int dr rJ_2(pr) \frac{1}{-r^2\ln\frac{1}{(|\Lambda_{\rm QCD}r)^2}} \left[ e^{2 Q_s^2\hat \Gamma(r)}-1\right]\,.\\
\end{split}
\end{equation}

\subsection{The $\langle A_iA_jUUUU\rangle $ correlator}
As discussed in the text we approximate the  correlator by the factorized expression
\begin{equation}
\begin{split}
&\Big\langle  A^i_a(\mathbf{R})A^{i'}_b(\mathbf{R}')  \left[U(\mathbf{z}_3)U^{\dagger}(R)\right]^{ca}
\left[U(\mathbf{z}'_3)U^{\dagger}(\mathbf{R}')\right]^{cb} \Big\rangle\\
\simeq &\Big\langle  A^i_a(\mathbf{R})A^{i'}_b(\mathbf{R}')\Big\rangle \Big\langle   \left[U(\mathbf{z}_3)U^{\dagger}(R)\right]^{ca}
\left[U(\mathbf{z}'_3)U^{\dagger}(\mathbf{R}')\right]^{cb} \Big\rangle\\
&+ \Big\langle  A^i_a(\mathbf{R})  \left[U(\mathbf{z}_3)U^{\dagger}(R)\right]^{ca}\Big\rangle \Big\langle
A^{i'}_b(\mathbf{R}')\left[U(\mathbf{z}'_3)U^{\dagger}(\mathbf{R}')\right]^{cb} \Big\rangle\\
&+\Big\langle  A^i_a(\mathbf{R})\left[U(\mathbf{z}'_3)U^{\dagger}(\mathbf{R}')\right]^{cb}\Big\rangle \Big\langle A^{i'}_b(\mathbf{R}')  \left[U(\mathbf{z}_3)U^{\dagger}(R)\right]^{ca}
 \Big\rangle.\\
\end{split}
\end{equation}
We focus on the last term and compute the following expectation value (we use the approach of ref.~\cite{Kovner:2001vi}) 
\begin{equation}
\begin{split}
&\Big \langle  A^i_a(\mathbf{R})\left[U(\mathbf{z}'_3)U^{\dagger}(\mathbf{R}')\right]^{cb}\Big\rangle\\
=&-\int_{-\infty}^{\infty} dx^- \Big\langle U^{ad}(x^-, \mathbf{R})\partial^i_{\mathbf{R}}\tilde{A}^+_d(x^-, \mathbf{R}) U^{ce} (\mathbf{z}'_3)U^{be}(\mathbf{R}')\Big\rangle\\
=&-\int_{-\infty}^{\infty} dx^-\int_{-\infty}^{\infty} dz^-  \Big\langle U^{ad}(x^-, \mathbf{R}) U^{cc'} (-\infty, z^-; \mathbf{z}'_3) T^m_{c'e'}U^{e'e}(z^-, +\infty; \mathbf{z}'_3) U^{be}(\mathbf{R}')\Big\rangle\\
&\qquad \times \left\langle  \partial^i_{\mathbf{R}}\tilde{A}^+_d(x^-, \mathbf{R}) ig\tilde{A}^+_m(z^-, \mathbf{z}'_3) \right \rangle  +  (c\leftrightarrow b, \mathbf{z}'_3 \leftrightarrow \mathbf{R}')\,.
\end{split}
\end{equation}
The correlator of two Wilson lines  is proportional to the dipole  in the adjoint representation 
\begin{equation}
\begin{split}
\Big\langle U^{e'e}(x^-, +\infty; \mathbf{z}'_3) U^{b'e}(x^-, +\infty; \mathbf{R}')\Big\rangle
=&\delta^{e'b'}D_g(x^-, +\infty; \mathbf{R}', \mathbf{z}'_3).
\end{split} 
\end{equation}
The correlator of three Wilson lines  can be readily computed  
\begin{equation}
\begin{split}
&\left\langle U^{ad}(-\infty, x^-; \mathbf{R}) U^{cc'} (-\infty, x^-; \mathbf{z}'_3)U^{bb'}(-\infty, x^- ; \mathbf{R}')\right\rangle  T^d_{c'b'}\\
=& T^a_{cb} \mathrm{exp}\left\{ \frac{N_c}{4} g^4\int^{x^-}_{-\infty} dz^- \mu^2(z^-) (\Gamma(\mathbf{R}, \mathbf{R}') + \Gamma(\mathbf{R}, \mathbf{z}'_3) + \Gamma(\mathbf{z}'_3, \mathbf{R}'))\right\}. 
\end{split} 
\end{equation}

Combining the ingredients, we obtained 
\begin{equation}\label{eq:AUU_one}
\begin{split}
&\left \langle  A^i_a(\mathbf{R})\left[U(\mathbf{z}'_3)U^{\dagger}(\mathbf{R}')\right]^{cb}\right\rangle\\
=&T^a_{cb}\left[  e^{\frac{1}{2}N_cg^4\bar{\mu}^2\Gamma(\mathbf{R}', \mathbf{z}'_3)} \right]  \frac{(-ig)\partial^i_{\mathbf{R}}[L(\mathbf{R}, \mathbf{z}'_3) - L(\mathbf{R}, \mathbf{R}')] }{\frac{N_c}{4}g^2 (\Gamma(\mathbf{R}, \mathbf{R}') + \Gamma(\mathbf{R}, \mathbf{z}'_3) - \Gamma(\mathbf{z}'_3, \mathbf{R}'))}  \left[ e^{ \frac{N_c}{4} g^4\bar{\mu}^2 (\Gamma(\mathbf{R}, \mathbf{R}') + \Gamma(\mathbf{R}, \mathbf{z}'_3) - \Gamma(\mathbf{z}'_3, \mathbf{R}'))} -1\right].  \\
\end{split}
\end{equation}
Similarly, analysis also gives 
\begin{equation}\label{eq:AUU_two}
\begin{split}
& \left\langle A^{i'}_b(\mathbf{R}')  \left[U(\mathbf{z}_3)U^{\dagger}(\mathbf{R})\right]^{ca}
 \right\rangle\\
 =&T^b_{ca}\left[  e^{\frac{1}{2}N_cg^4\bar{\mu}^2\Gamma(\mathbf{R}, \mathbf{z}_3)} \right]  \frac{(-ig)\partial^{i'}_{\mathbf{R}'}[L(\mathbf{R}', \mathbf{z}_3) - L(\mathbf{R}', \mathbf{R})] }{\frac{N_c}{4}g^2 (\Gamma(\mathbf{R}', \mathbf{R}) + \Gamma(\mathbf{R}', \mathbf{z}_3) - \Gamma(\mathbf{z}_3, \mathbf{R}))}  \left[ e^{ \frac{N_c}{4} g^4\bar{\mu}^2 (\Gamma(\mathbf{R}', \mathbf{R}) + \Gamma(\mathbf{R}', \mathbf{z}_3) - \Gamma(\mathbf{z}_3, \mathbf{R}))} -1\right].  \\
 \end{split}
\end{equation}
As a consistency check on our expression,  consider  the limit $\mathbf{R} \rightarrow \mathbf{z}_3$, when 
$\left\langle A^{i'}_b(\mathbf{R}')  \left[U(\mathbf{z}_3)U^{\dagger}(\mathbf{R})\right]^{ca}
 \right\rangle = \langle A^{i'}_b(\mathbf{R}') A^j_e(\mathbf{z}_3)\rangle  T^e_{ca} (-ig) (\mathbf{R}-\mathbf{z}_3)^j$. Indeed this can be explicitly reproduced from our final expression~\eqref{eq:AUU_one}:  
\begin{equation}
\begin{split}
& \left\langle A^{i'}_b(\mathbf{R}')  \left[U(\mathbf{z}_3)U^{\dagger}(\mathbf{R})\right]^{ca}
 \right\rangle  \\
 =&T^b_{ca} \frac{(-ig)(\mathbf{R}-\mathbf{z}_3)^j\partial^{i'}_{\mathbf{R}'}[\partial^j_{\mathbf{z}_3}L(\mathbf{R}', \mathbf{z}_3) ] }{\frac{N_c}{2}g^2 \Gamma(\mathbf{R}', \mathbf{z}_3) }  \left[ e^{ \frac{N_c}{2} g^4\bar{\mu}^2 \Gamma(\mathbf{R}', \mathbf{z}_3) } -1\right]\\
  =& \langle A^{i'}_b(\mathbf{R}') A^j_e(\mathbf{z}_3)\rangle  T^e_{ca} (-ig) (\mathbf{R}-\mathbf{z}_3)^j.  \\
 \end{split}
\end{equation}

In the limit  $|\mathbf{R} -\mathbf{R}'|\lesssim 1/p_3\ll 1/Q_s$, eqs. \eqref{eq:AUU_one} and \eqref{eq:AUU_two} can be further simplified by noting that the exponential factor  can be expanded due to the smallness of the expression in the second exponent, $Q_s^2 (\Gamma(\mathbf{R}, \mathbf{R}') + \Gamma(\mathbf{R}, \mathbf{z}'_3) - \Gamma(\mathbf{z}'_3, \mathbf{R}')) \ll 1$. We keep only the first-non-trivial order to get 
\begin{equation}
\begin{split}
&\left \langle  A^i_a(\mathbf{R})\left[U(\mathbf{z}'_3)U^{\dagger}(\mathbf{R}')\right]^{cb}\right\rangle
\approx T^a_{cb}\left[  e^{2 Q_s^2\hat \Gamma(\mathbf{R}', \mathbf{z}'_3)} \right]  (-ig)\partial^i_{\mathbf{R}}[L(\mathbf{R}, \mathbf{z}'_3) - L(\mathbf{R}, \mathbf{R}')] \frac{4\pi Q_s^2}{N_c g^2}     \\
\end{split}
\end{equation}
and
\begin{equation}\label{eq:AUU_two}
\begin{split}
& \left\langle A^{i'}_b(\mathbf{R}')  \left[U(\mathbf{z}_3)U^{\dagger}(\mathbf{R})\right]^{ca}
 \right\rangle
 \approx T^b_{ca}\left[  e^{2 Q_s^2 \hat \Gamma(\mathbf{R}, \mathbf{z}_3)} \right]  (-ig)\partial^{i'}_{\mathbf{R}'}[L(\mathbf{R}', \mathbf{z}_3) - L(\mathbf{R}', \mathbf{R})] \frac{4 \pi Q_s^2}{N_c g^2}  .  \\
 \end{split}
\end{equation}

\bibliography{trijet_dense_draft}

\end{document}